\documentclass[prc,aps,fleqn,twocolumn,twoside,nofootinbib]{revtex4}

\usepackage{latexsym,exscale}
\usepackage{epsfig}
\usepackage{rotating}
\usepackage{amsmath,amssymb,amsfonts}
\usepackage{graphicx}

\begin{document}

\title{
Hydrodynamical evolution near the QCD critical end point}
\author{Chiho Nonaka}
\affiliation{Department of Physics, Duke University,
             Durham, NC 27708, U.S.A.}
\author{Masayuki Asakawa}
\affiliation{Department of Physics, Osaka University, Toyonaka 560-0043, Japan}
\date{\today}

\begin{abstract}
Hydrodynamical calculations have been successful in describing 
global observables in ultrarelativistic heavy ion collisions,
which aim to observe the production of the quark-gluon plasma.
On the other hand, recently, a lot of evidence that there exists
a critical end point (CEP) in the QCD phase diagram has been accumulating.
Nevertheless, so far, no equation of state with the CEP has been
employed in hydrodynamical calculations. In this paper, we construct
the equation of state with the CEP on the basis of the universality
hypothesis and show that the CEP acts as an attractor
of isentropic trajectories. We also consider the time evolution
in the case with the CEP and discuss how the CEP affects the final
state observables, such as the correlation length, fluctuation, 
chemical freezeout, kinetic freezeout, and so on. Finally,
we argue that the anomalously low kinetic freezeout temperature
at the BNL Relativistic Heavy Ion Collider suggests the possibility of 
the existence of the CEP. 
\end{abstract}

\maketitle

\section{Introduction}
The structure of the QCD phase diagram is one of the most interesting
and important topics in nuclear and particle physics,
and a lot of intensive studies have been carried out
from both theoretical side and experimental side \cite{QM04}.
Recent theoretical studies have revealed the rich structure in
the QCD phase diagram at finite density, which consists of not
only the quark-gluon plasma (QGP) phase and hadron phase but also
the two flavor color superconductor phase and color-flavor locked phase
\cite{RaWi}.
Furthermore, in the finite temperature direction,
recent lattice calculations suggest the existence of
mesonic bound states even above the deconfinement phase transition
temperature \cite{AsHaNa03,AsHa03,Bielefeld02,DaKaPeWe03,UmNoMa02,MrLeRhSh}.

In this paper, we focus on the critical end point (CEP) in the
QCD phase diagram,
which is the terminating point of the first order phase transition
\footnote
{The terminology `critical end point' is sometimes used for
a different meaning \cite{FiBa91}. However, it is rather
customary in hadron physics to use it for the meaning we employ
here. Thus, we use `critical end point' instead of 'critical point'
in this paper. We thank M. Stephanov for informing us of Ref.\ \cite{FiBa91}.
}.
In particular, we will consider its effects on the
observables in ultrarelativistic heavy ion collisions in detail.
Several effective theory analyses have predicted the existence
of the CEP in the QCD phase diagram \cite{AsaYa89,random,St}. 
Since Fodor and Katz presented a QCD phase diagram
with the CEP for the first time with
lattice calculation \cite{FoKa02},  a lot of remarkable progress
in finite temperature and density lattice calculation
has been made \cite{Swansea02,FoPh}. 
However, unfortunately, it will need more time before we reach
the conclusive result on the existence of the CEP, its precise position,
and so on, because finite density lattice QCD still has difficulties in 
performing accurate calculation on large lattices
and at large chemical potential.
Here, the important point is that the existence of the 
CEP was shown semi-quantitatively by non-perturbative
QCD calculation. The precise determination
of the location of the CEP in the QCD phase diagram
is a difficult problem. The locations of the CEP in different studies are,
in fact, scattered over
the $T$ (temperature)-$\mu_B$ (baryon chemical potential)
plane \cite{St}. Therefore, analyses from the experimental
side are also indispensable 
in order to understand the properties of the QCD critical end point. 

A promising way to investigate the existence and location of the
QCD critical end point is to carry out experiments at various
collision energies and compare their results.
However, we have to find the observables which are suitable for  
detecting the CEP \cite{StRaSh,HaIke02,Fu02,HaSte03,ScMoMiRi00}
for that purpose. In this paper, we study the CEP from the point of view of 
phenomenological analyses which can be compared with experimental 
results easily and discuss the indication of the existence of the CEP
in experimental observables. 

We divide the discussion into three steps.
First, we construct equations of state that include 
the CEP. This step is necessary in order to consider the hydrodynamical
evolution of the system. The guiding principle here is
the theory of critical phenomena, in particular, the universality
hypothesis. Second, we consider the deviation from the thermal equilibrium
near the CEP. This is important because i) the typical time scale
is elongated near the CEP, and ii) the system produced in
ultrarelativistic heavy ion collisions cannot stay near
the CEP due to the strong expansion. Finally, we discuss
the possibility to observe the consequences of the CEP experimentally
by using the preceding results.

The structure of this paper is as follows. In Section II 
we argue the general feature of the equation of state with the CEP
on the basis of the universality hypothesis,
the focusing effect in isentropic trajectories 
in the $T$-$\mu_B$ plane, and hydrodynamical expansion
near the CEP.      
In Section III we investigate the growth and decrease of the correlation
length near the CEP from the point of view of slowing-out-of-equilibrium. 
In Section IV we discuss the consequences of the CEP 
in ultrarelativistic heavy ion collision experiments.
Section V is devoted to summary. 

\section{Equation of State with the Critical End Point}

In this section, we construct realistic equations of state including 
the CEP. 
The equation of state with the CEP consists of two parts, 
the singular part and non-singular part. We assume that the CEP
in QCD belongs to the same universality class as that in
the three dimensional Ising model on the basis of the
universality hypothesis. After mapping the variables and 
the equation of state near
the CEP in the three dimensional Ising model onto those in QCD,
we match the singular entropy density near the 
CEP with the non-singular QGP and hadron phase
entropy densities which are known away from the CEP.  
From this procedure we determine the behavior of the entropy density 
which includes both the singular part and non-singular part   
in a large region in the $T$-$\mu_B$ plane. 
Finally, from the entropy density, we extract the behavior
of other thermodynamical quantities such as the baryon number density,
pressure, and energy density, and show that in the
$T$-$\mu_B$ plane the CEP acts as
an attractor of isentropic trajectories, $n_B/s = {\rm const.}$,
with $n_B$ and $s$ being the baryon number density and entropy density,
respectively.

\subsection{Singular Part of Equation of State}

In the three dimensional Ising model, the magnetization $M$ 
(the order parameter) 
is a function of the reduced temperature $r=(T-T_c)/T_c$ and  
the external magnetic field $h$ with $T_c$ being
the critical temperature.
The CEP is located at the origin
$(r,h)=(0,0)$.
At $r<0$ the order of the
phase transition is first and at $r>0$
it is crossover.
A useful form of the equation of state of the three-dimensional Ising model 
is given through the parametric representation by the two variables
$R$ and $\theta$,  
\begin{eqnarray}
\left \{
\begin{array}{lcl}
M & = & M_0 R^\beta \theta,  \\
h & = & h_0 R^{\beta\delta} \tilde{h}(\theta) \\
  & = & h_0 R^{\beta \delta}
(\theta - 0.76201 \theta ^3 + 0.00804 \theta^5), \\ 
r & = & R(1-\theta^2) \hspace{0.5cm}( R \geq 0, -1.154 \leq \theta \leq 1.154),
\end{array}
\right .
\label{Eq-3deos}
\end{eqnarray}
where $M_0$ and  $h_0$ are normalization constants and critical exponents 
$\beta$ and $\delta$ are 0.326 and  4.80, respectively \cite{GuZi}. 
We set the normalization constants $M_0$ and $h_0$ by imposing     
$M(r\!\!=\!\!-1, h\!\!=\!\!+0)=1$ and  $M(r\!\!=\!\! 0, h\!\!=\!\!1)=1$, 
which assures that the critical 
magnetization behaves as
$M(r\!\!=\!\!0, h) \propto {\rm sgn}(h) |h|^{1/\delta}$ 
and $M(r,h=+0) \propto |r|^{\beta}$
$ (r < 0 )$ around the origin. 

In order to determine the singular part of the entropy density, 
we start from the Gibbs free energy density $G(h,r)$,   
\begin{equation}
G(h,r) = F(M,r)-Mh,
\end{equation}
where $F(M,r)$ is the free energy density.
Setting the free energy density,
\begin{equation}
F(M,r)  =  h_0 M_0 r^{2-\alpha} g(\theta),  
\label{Eq-F}
\end{equation}
together with the relation $h = (\partial F/ \partial M)_r$, 
we obtain the differential equation for $g(\theta)$,
\begin{equation}
\tilde{h}(\theta)(1-\theta^2 + 2 \beta \theta^2) = 
2 (2- \alpha) \theta g( \theta) + (1 - \theta^2) g'( \theta), 
\label{Eq-diff}
\end{equation}
where a critical exponent $\alpha$ is 0.11 \cite{GuZi}.  
Substituting $\tilde{h}(\theta)$ into Eq.\ (\ref{Eq-diff}), 
we obtain
\begin{equation}
g(\theta)=g(1) + c_1(1-\theta^2)+c_2(1-\theta^2)^2 + c_3 (1-\theta^2)^3, 
\end{equation}
where $g(1)$ is determined by Eqs.\ (\ref{Eq-3deos}) and (\ref{Eq-diff}), 
\begin{equation}
g(1) = \frac{\beta}{2 - \alpha}\tilde{h}(1),
\end{equation}
and the coefficients $c_1$, $c_2$, and $c_3$ are given by  
\begin{eqnarray}
c_1 & = & - \frac{1}{2} \frac{1}{\alpha-1} \left \{
(1 - 2 \beta)(1+a+b)-2 \beta(a + 2b)
\right \},   \nonumber \\
c_2 & = &
 - \frac{1}{2 \alpha} \left \{
 2 \beta b - (1- 2 \beta) (a+2b)
 \right \},  \nonumber \\  
c_3 & = & - \frac{1}{2 (\alpha + 1)}b(1-2\beta), 
\end{eqnarray}
where $a=-0.76201$ and $b=0.00804$ are the coefficients of
the $\theta^3$ and $\theta^5$ terms in $\tilde{h}(\theta)$,
respectively \cite{GuZi}.
Actually, there is freedom to add the function
$C(1-\theta^2)^{2-\alpha}$ to $g(\theta )$,
with $C$ being a constant.
However, we will not take into account this part with a free parameter,
because, whereas the universality tells us about only the critical behavior
near the phase transition, this part does not show a singular
behavior.

Differentiating the Gibbs free energy by the temperature, 
we obtain the singular part of the entropy density $s_c$ near the QCD critical
end point,
\begin{eqnarray}
s_c & = & - \left ( \frac{\partial G}{\partial T} \right )_{\mu_B}  \nonumber \\
    & = & - \left ( \frac{\partial G}{\partial h} \right )_r
\frac{\partial h}{\partial T} 
- \left ( \frac{\partial G}{\partial r} \right )_h
\frac{\partial r}{\partial T},
\label{Eq-Sc}
\end{eqnarray}
where $\left ( \frac{\partial G}{\partial h } \right )_r$ and  
$\left ( \frac{\partial G}{\partial r } \right )_h$
are, respectively, given by
\begin{eqnarray}
\left ( \frac{\partial G}{\partial h}\right )_r  
& = & \left ( \frac{\partial F(M,r)}{\partial h} \right )_r
 -\left ( \frac{\partial M}{\partial h} \right )_r h -M \nonumber \\
& = & -M, 
\label{Eq-S1}
\end{eqnarray}
\begin{widetext}
\begin{eqnarray}
\left (
\frac{\partial G}{\partial r}
\right )_h & = & 
\left ( 
\frac{\partial F(M,r)}{\partial r}
\right )_h - 
\left (
\frac{\partial M}{\partial r}
\right )_h h \nonumber \\
 & = & \frac{h_0M_0R^{1-\alpha}}{
2 \beta \delta \theta \tilde{h}(\theta)
+  (1-\theta^2) \tilde{h}'(\theta)} 
\times 
\left \{  
(2-\alpha)\tilde{h'}(\theta)g(\theta) - \beta \theta \tilde{h'}(\theta)
\tilde{h}(\theta) - \beta \delta \tilde{h}(\theta)g'(\theta) 
+ \beta \delta \tilde{h}^2(\theta)
\right \}.
\label{Eq-S2}
\end{eqnarray}
\end{widetext}
Note that $T$ in Eq.\ (\ref{Eq-Sc}) is the temperature on
the QCD side. 
In Eqs.\ (\ref{Eq-S1}) and (\ref{Eq-S2}),
we use the differential equation Eq.\ (\ref{Eq-diff}) and 
the following relations,
\begin{eqnarray}
\left (
\frac{\partial R}{\partial h}
\right )_r&  = &  
\frac{1}{h_0 R^{\beta \delta-1}} \frac{2 \theta}
{2 \beta \delta \theta \tilde{h}(\theta) 
+ (1-\theta^2)\tilde{h'}(\theta)}, \nonumber \\
\left (
\frac{\partial \theta}{\partial h}
\right )_ r & = & 
\frac{1}{h_0 R^{\beta \delta}} \frac{1 - \theta^2}
{ 2 \beta \delta \theta \tilde{h} ( \theta ) + 
(1-\theta^2) \tilde{h'}(\theta)}, \nonumber \\
\left (
\frac{\partial R}{\partial r}
\right )_h & = & 
\frac{\tilde{h'}(\theta)}
{ 2 \beta \delta \theta \tilde{h} ( \theta ) + (1-\theta^2) \tilde{h'}(\theta) 
}, \nonumber \\ 
\left (
\frac{\partial \theta}{\partial r}
\right )_h & = &  
-\frac{\beta \delta}{R}\frac{\tilde{h}(\theta)}
{2  \beta \delta \theta \tilde{h}(\theta)+ (1-\theta^2)\tilde{h'}(\theta)}.
\end{eqnarray}

\begin{figure}
\includegraphics[width=0.8\linewidth]{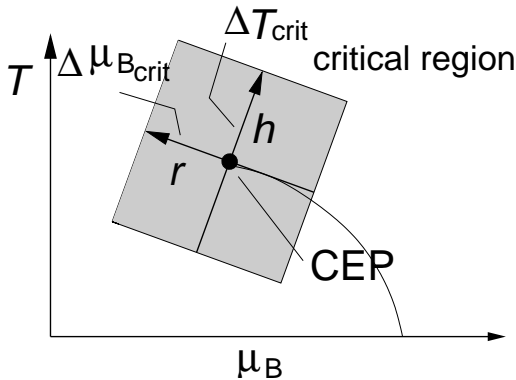}
\caption{\label{Fig-map}
Sketch of the $r$-$h$ axes (three dimensional Ising model)
mapped onto the $T$-$\mu_{B}$ plane (QCD).  
The $r$ axis is tangential to the QCD phase
boundary at the CEP. We set the $h$ direction
perpendicular to the $r$ direction.
}
\end{figure}

Now we map the $r$-$h$ plane in the three dimensional Ising model 
onto the $T$-$\mu_B$ plane in QCD in order to determine 
$\partial h/\partial T$ and 
$\partial r / \partial T$ in Eq.\ (\ref{Eq-Sc}). 
The CEP in the three dimensional Ising Model,
which is the origin in the $r$-$h$ plane, is mapped to the CEP in QCD,
$(T,\mu_B) = (T_E,\mu_{BE})$.
The $r$ axis is tangential to the first order phase transition
line at the CEP \cite{GeRt}. 
However, there is no general rule about how the $h$ axis is
mapped in the $T$-$\mu_B$ plane. 
Here, for simplicity, we set the $h$ axis perpendicular to the $r$ axis.
In Fig.\ \ref{Fig-map}, at $r<0$ the 
order of the phase transition is first and 
at $r>0$ it is crossover.

\subsection{Thermodynamical Quantities}
\begin{figure}[h]
\begin{minipage}{0.49\linewidth}
\includegraphics[width=1.05\linewidth]{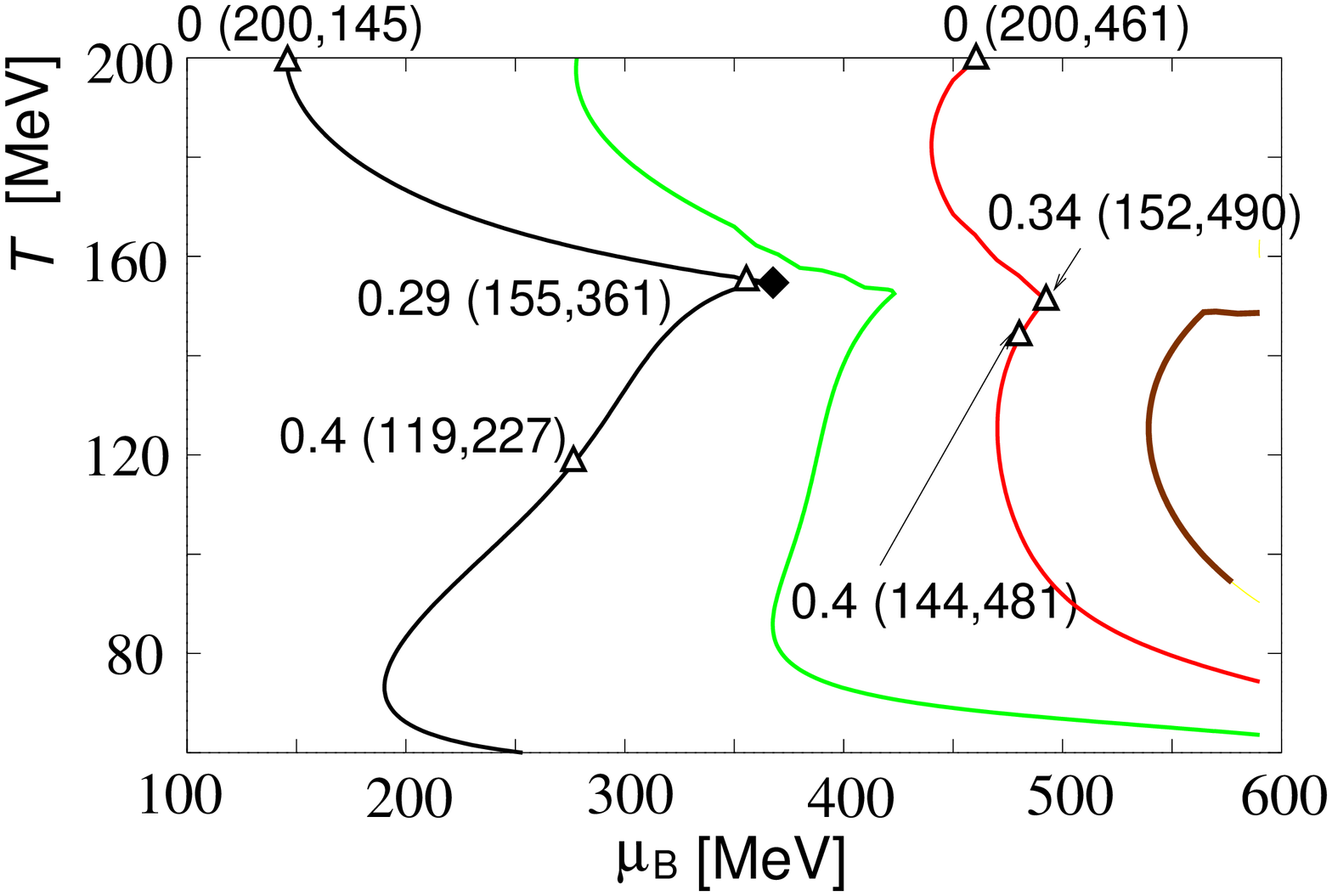}
\end{minipage}
\begin{minipage}{0.49\linewidth}
\includegraphics[width=1.05\linewidth]{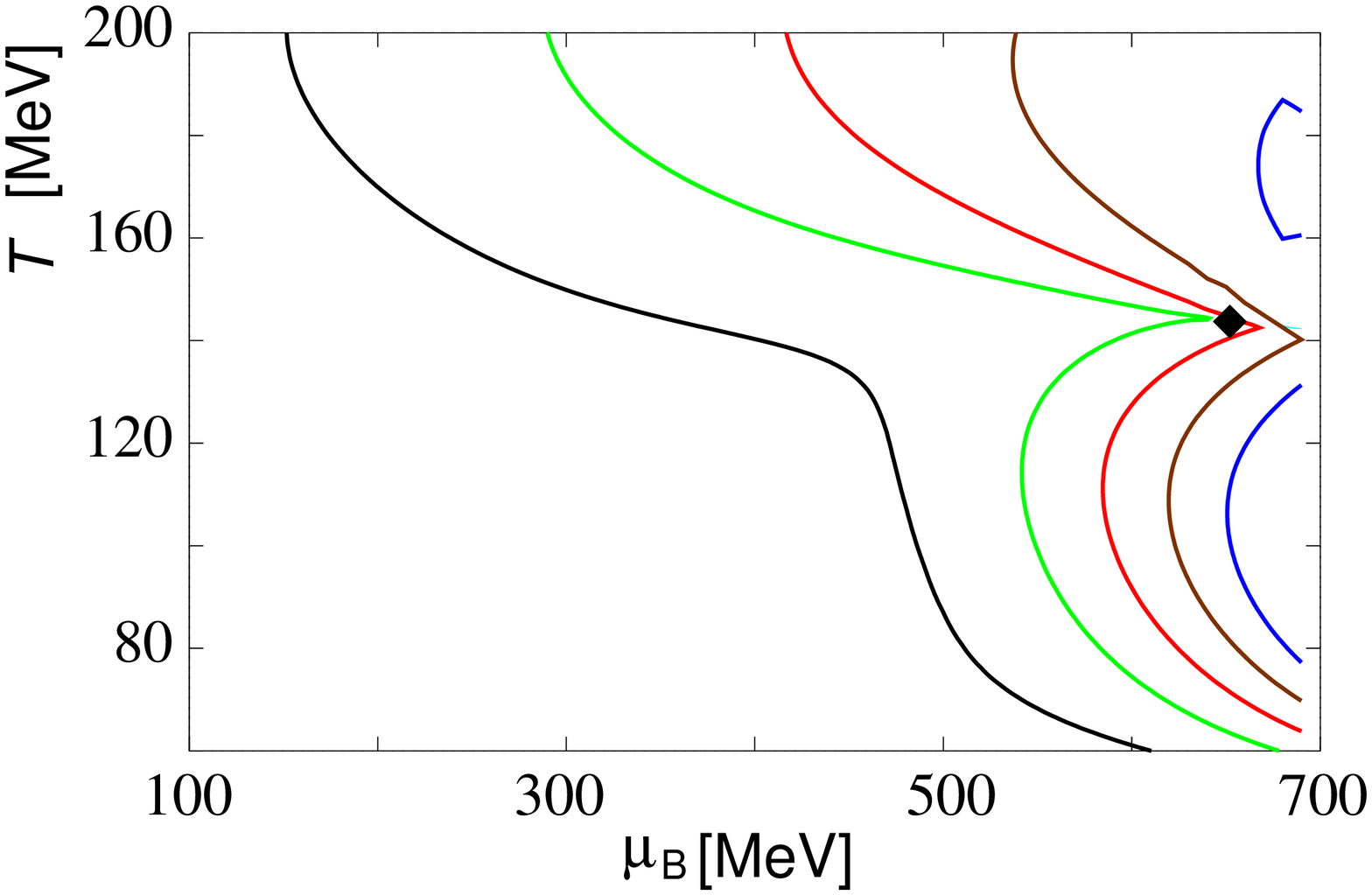}
\end{minipage}
\caption{\label{Fig-nbs}
Isentropic trajectories in the cases with the CEP.
The CEP is located at $(T_{\rm E}, \mu_{B{\rm E}}) = 
(154.7 \hspace{2mm}{\rm MeV}, 
~367.8 \hspace{2mm}{\rm MeV})$ (left) and   
$(T_{\rm E}, \mu_{B{\rm E}}) = (143.7 \hspace{2mm}{\rm MeV}, 
~652.0 \hspace{2mm}{\rm MeV})$ (right).
The values of $n_B/s$ on the trajectories are 
0.01, 0.02, 0.03, and 0.04 (left) and   
0.01, 0.02, 0.03, 0.04, and 0.05 (right) from left to right.
$L/L_{\rm total} (T,\mu_B )$ is
shown for some points on two trajectories in the left figure.
}
\end{figure}

For quantitative construction of equations of state
with the CEP, 
we fix the relation between the scales in
$(r,h)$ and $(T,\mu_B)$ variables,
which provides the size of the critical region
around the CEP in the $T$-$\mu_B$ plane, as follows:
$\Delta r = 1$ ($r$-$h$ plane) $\leftrightarrow \Delta \mu_{B\rm crit}$ 
($T$-$\mu_B$ plane) and  
$\Delta h = 1$ ($r$-$h$ plane) $\leftrightarrow 
\Delta T_{\rm crit}$ ($T$-$\mu_B$ plane) (Fig.\ \ref{Fig-map}).
$\Delta\mu_{B{\rm crit}}$ and $\Delta T_{\rm crit}$ give,
respectively, approximate extensions of the critical region
in the $\mu_{B}$ and $T$ directions when the location of the CEP
is close to the $T$ axis and, as a result, the $r$ axis is approximately
parallel to the $\mu_{B}$ axis, as recent lattice calculations suggest
\cite{FoKa02,Swansea02,FoPh}.
In order to connect the equations of state in the singular
region and the non-singular region smoothly,
we define the dimensionless variable $S_c(T, \mu_B)$ for the singular
part of the entropy density $s_c$, which has the dimension [energy]$^{-1}$,
\begin{equation}
S_c(T,\mu_B) = A(\Delta T_{\rm crit}, \Delta \mu_{B\rm crit})
s_c(T,\mu_B),
\label{Eq-size}
\end{equation}
where $A(\Delta T_{\rm crit}, \Delta \mu_{B\rm crit})
=\sqrt{\Delta T^2_{\rm crit}  + \Delta \mu^2_{B\rm crit}} \times D$ 
and $D$ is a dimensionless constant.
The extension of the critical domain around the CEP is specified
by the parameters $\Delta T_{\rm crit}$,
$\Delta \mu_{B\rm crit}$, and $D$. 
\begin{figure}
\includegraphics[width=0.9\linewidth]{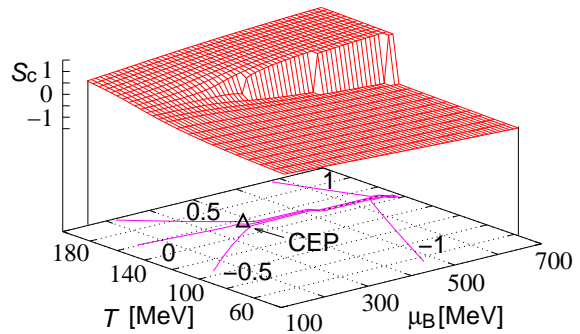}
\caption{\label{Fig-Scrit}
Dimensionless variable $S_c(T,\mu_B)$ as a function of $T$ and 
$\mu_B$. The CEP is located at
$(T,\mu_B) =
(154.7 \hspace*{2mm}{\rm MeV}, 367.8 \hspace*{2mm}{\rm MeV})$
(the triangle in the contour plot).
}
\end{figure}
In Fig.\ \ref{Fig-Scrit}, an example of $S_c$ is shown 
as a function of $T$
and $\mu_B$.
The CEP is located at 
$(T,\mu_B)=$(157.4 MeV, 367.8 MeV) (the triangle in the
contour plot). 
We use the same parameters $(\Delta T_{\rm crit}, \Delta 
\mu_{B {\rm crit}}, D)$ as those  
in Fig. \ref{Fig-nbs} (left). $S_c$ is obtained by solving numerically
the implicit relations derived in the preceding subsection.
It is clearly seen that 
the order of the phase transition changes from crossover to first order at
the CEP. The contour lines, $S_c = \pm 0.5$, give a rough idea about
the size of the phase transition region. At lower $\mu_B$ below the
critical chemical potential, the phase transition is smeared and
the effect of the phase transition is observed in a larger domain
in the temperature direction. Using the dimensionless variable
$S_c(T, \mu_B)$,
we define the entropy density in the $T$-$\mu_B$ plane,
\begin{widetext}
\begin{equation}
s (T, \mu_B)  =  \frac{1}{2}
\left( 1 - \tanh [S_c (T, \mu_B) ]\right ) s_{\rm H}(T, \mu_B) 
 + \frac{1}{2}
\left( 1 + \tanh[S_c (T, \mu_B)] \right ) s_{\rm Q}(T, \mu_B), 
\label{Eq-rentro}
\end{equation}
\end{widetext}
where $s_{\rm H}$ and $s_{\rm Q}$ are the entropy densities in the
hadron phase and QGP phase away from the CEP, respectively.
This entropy density includes both singular and non-singular
contributions, and more importantly, gives the correct
critical exponents near the QCD critical end point.
$s_{\rm H}$ is calculated from the equation of state
of the hadron phase in the excluded volume approximation \cite{RiGo}, 
\begin{eqnarray}
P(T, \{\mu_{Bi}\}) & = & \sum_i P^{\rm ideal}_i
(T, \mu_{Bi} - V_0 P(T, \{\mu_{Bi} \})) \nonumber \\ 
& = & \sum_i P^{\rm ideal}(T, \tilde{\mu}_{Bi}), 
\end{eqnarray}
where $P$ is the pressure, $P^{\rm ideal}_i$ is
the pressure of the ideal gas of particle $i$,
$\mu_{Bi}$ is the baryon chemical potential of particle $i$, 
and $V_0$ is the hadron volume common to all hadron species.
Non-strange resonances with mass up to 2 GeV
are included in the sum 
and the radius of hadrons is fixed at 0.7 fm.
We obtain $s_{\rm Q}$ from 
the equation of state of the QGP phase in the Bag model,
\begin{eqnarray}
P(T,\mu_B) & = & \frac{(32 + 21N_f)\pi^2}{180}T^4 + \frac{N_f}{2}
\left ( \frac{\mu_B}{3}  
\right )^2 T^2 \nonumber \\
 & &+  
\frac{N_f}{4 \pi^2} 
\left (  \frac{\mu_B}{3}    \right )^4 -B, 
\end{eqnarray}
where the number of the flavors $N_f$ is 2 and the bag constant $B$ is 
$(220 \hspace{1mm}{\rm MeV})^4$. 
In Eq.\ (\ref{Eq-rentro}) the relative strength of the 
singularity also depends on the 
distance between the QCD critical end point, where $S_c =0$,
and the $s_{\rm Q}=s_{\rm H}$ line. 
If the CEP is located near the $s_{\rm Q}=s_{\rm H}$ line, less singularity is realized.  
However, in the following calculations, we assume that the CEP is on
the phase transition line in the bag plus excluded volume model,
where the phase transition is always of strong first order as shown later.
Thus, $s_{\rm Q}$ is substantially larger than $s_{\rm H}$ at the CEP and
the singularity is not suppressed.

Once we construct the entropy density,
we can calculate the other thermodynamical quantities
such as the baryon number density, pressure, and energy density.
The baryon number density $n_B$ is given by
\begin{eqnarray}
n_B(T,\mu_B)
& = & \frac{\partial P }{\partial \mu_B} \nonumber \\
& = & \int_{0}^{T} \frac{\partial s (T', \mu_B) }{\partial \mu_B}dT' + 
n_B(0,\mu_B).
\label{Eq-nb}
\end{eqnarray}
In the first order phase transition region, in order to take into account
the discontinuity in the entropy density and baryon number density
on the phase boundary, it is necessary to add the following term to 
Eq.\ (\ref{Eq-nb}) for $T > T_c (\mu_B)$,
\begin{equation}
\left |
\frac{\partial T_c(\mu_B)}{\partial \mu_B}
\right |
(s(T_c(\mu_B)+0, \mu_B) -
s(T_c(\mu_B)-0, \mu_B)),
\end{equation}
where $T_c (\mu_B)$ is the temperature of the first order
phase transition at $\mu_B$.
The pressure $P$ is obtained by integrating the entropy density with
regard to the temperature,
\begin{equation}
P(T, \mu_B) = \int_{0}^{T}  s(T', \mu_B)dT' +
P(0,\mu_B).
\end{equation}
Finally, from the thermodynamical relation  
the energy density $E$ is obtained as
\begin{equation}
E = Ts - P - \mu_B n_B.
\end{equation}

Here we make a comment on the choice of the parameters
$\Delta T_{\rm crit}$, $\Delta \mu_{B\rm crit}$,
and $D$. The linear mapping from $(r,h)$ to $(T,\mu_B)$,
strictly speaking, holds only in the proximity of the CEP,
where the thermodynamical quantities comply with the universality.
We use the linear mapping throughout the critical region.
In order to avoid artifacts due to this simplification,
$\Delta T_{\rm crit}$, $\Delta \mu_{B\rm crit}$, and $D$
need to be appropriately chosen so that thermodynamical constraints
are not violated. In particular, the following thermodynamical
inequalities \cite{GeRt,LaLi,GlPr} need to be observed,
\begin{eqnarray}
\left(
\frac{\partial S}{\partial T}
\right)_{V,N_B}
& >  & 0, \nonumber \\
\left (
\frac{\partial P}{\partial V}
\right)_{T,N_B}
& >  & 0,  \nonumber \\
\left(
\frac{\partial \mu_B}{\partial N_B}
\right )_{T,V}
& >  & 0,     
\label{Eq-ctherm}
\end{eqnarray}
where $S$ is the entropy, $N_B$ is the baryon number, and
$V$ is the volume of the system.

\begin{figure}[h]
\begin{minipage}{0.49\linewidth}
\includegraphics[width=1.1\linewidth]{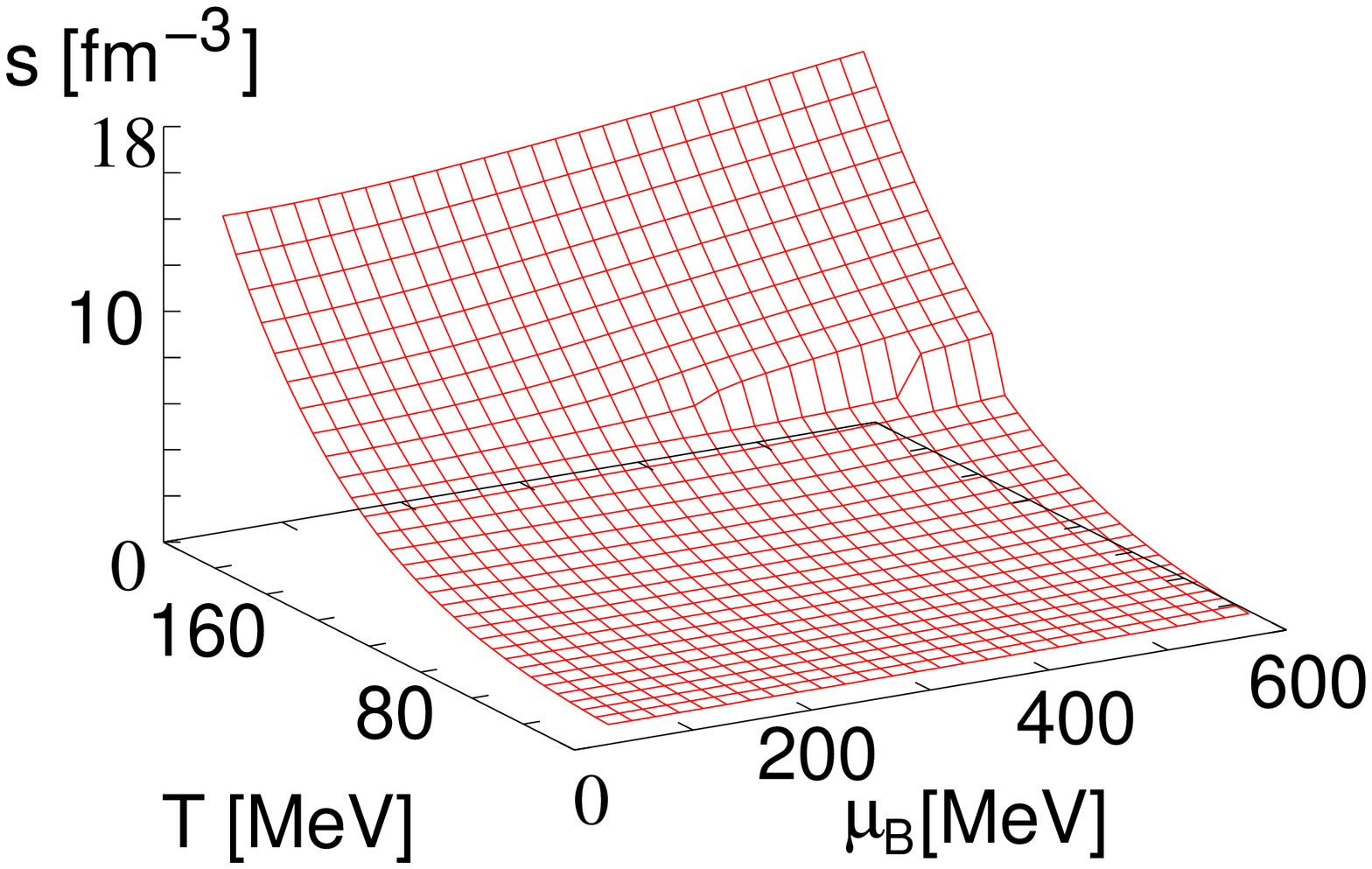}
\end{minipage}
\begin{minipage}{0.49\linewidth}
\includegraphics[width=1.1\linewidth]{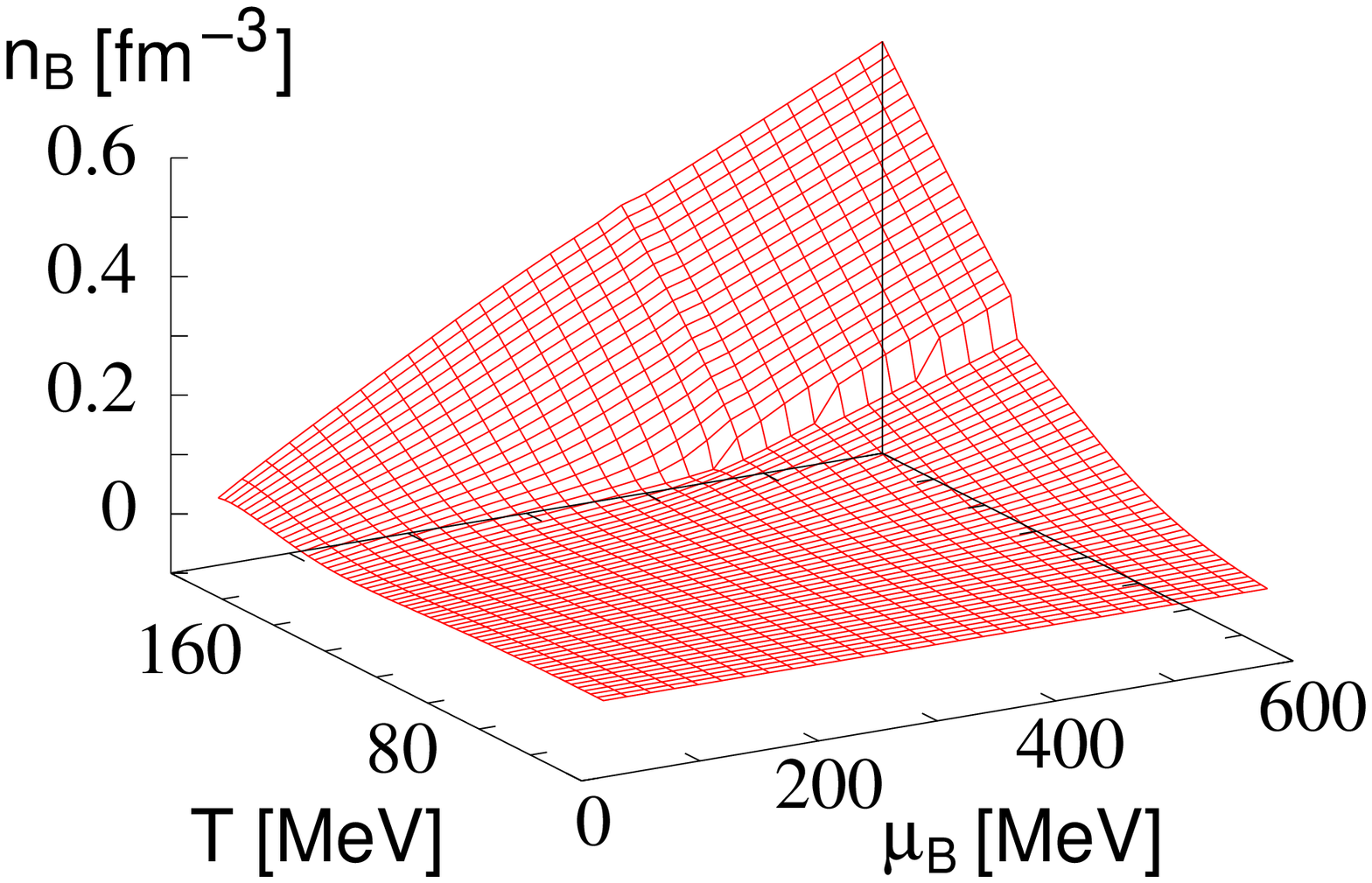}
\end{minipage}
\caption{\label{Fig-EOS}
Entropy density (left) and the 
baryon number density (right) as a function of $T$ and $\mu_B$.
$(T_{\rm E}, \mu_{B {\rm E}}) = (154.7 \hspace{2mm}{\rm  MeV}, 
~367.8 \hspace{2mm}{\rm  MeV})$. The values of
$(\Delta T_{\rm crit},\Delta \mu_{\rm Bcrit},D)$
are the same as in Fig.\ \ref{Fig-nbs} (left).
}
\end{figure}

Figure \ref{Fig-EOS} indicates the entropy density and 
the baryon number density as a function of $T$ and $\mu_B$.
We can see that in the high chemical potential region the order of the
phase transition is first and that in the low chemical potential region
a smooth phase transition occurs, which reflects the existence of the CEP. 
In this calculation the CEP is located at 
$(T, \mu_B)=(154.7 \hspace{2mm}{\rm  MeV},~367.8 \hspace{2mm}{\rm  MeV})$.
Since the universality does not provide the information on the
location of the CEP, it is treated as a parameter
throughout this paper as well as the values of  
$\Delta T_{\rm crit}$, $\Delta \mu_{B\rm crit}$, and $D$.
Note, however, that the local singular behavior of thermodynamical
quantities around the CEP is fixed by the universality
hypothesis and that accordingly the local features such as
the focusing, which we discuss below, are not affected by
this ambiguity.

First we investigate the behavior of the isentropic trajectories,
i.e., contour lines of $n_B/s$. When entropy production can be
ignored, the entropy and baryon number are conserved in each
volume element and, therefore, the temperature and
chemical potential in a given volume element change
along the contour lines specified by the initial condition.

\begin{figure}
\includegraphics[width=0.7\linewidth]{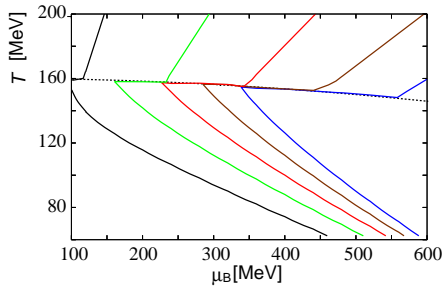}
\caption{\label{Fig-nbsc}
Isentropic trajectories (solid lines) in the bag plus excluded volume 
model. The values of $n_B/s$ on the trajectories are 
0.01, 0.02, 0.03, 0.04, and 0.05 from left to right.  
The dashed line stands for the phase boundary. 
The order of the phase transition is always first in this case.
}
\end{figure}

Figure \ref{Fig-nbs} shows the isentropic trajectories
in the $T$-$\mu_B$ plane.
The values of $(\Delta T_{\rm crit},~\Delta \mu_{B{\rm crit}},~D)$ for
the Fig.\ \ref{Fig-nbs} (left) and Fig.\ \ref{Fig-nbs} (right) are
$(100~{\rm MeV},~200~{\rm MeV},~0.15)$ and
$(100~{\rm MeV},~200~{\rm MeV},~0.4)$,
respectively. In both cases, the trajectories are focused
to the CEP. This `focusing' effect is more clearly observed
in the case of $(T_{\rm E}, \mu_{B{\rm E}})=(143.7 \hspace{2mm}{\rm  MeV}, 
~652 \hspace{2mm}{\rm  MeV})$ (Fig.\ \ref{Fig-nbs} (right)). 
Thus, the CEP acts as an attractor of isentropic trajectories. 
Figure \ref{Fig-nbsc} shows isentropic
trajectories in the bag plus excluded volume model,
which is currently employed in most of hydrodynamical 
calculations \cite{KoHe03}.
The order of the phase transition is always first in this case.
There is no focusing effect on
the isentropic trajectories in this case.
Instead, the trajectories are just shifted to the left
on the phase transition line.
This implies that the hydrodynamical
evolution in the case with the CEP
is very different from the one in the case with
the equation of state given by the bag plus excluded volume model. 
Because of this attractor character of the CEP,
it is not needed to fine-tune the collision energy
to make the system pass near the CEP, 
which is pointed out in Ref.\ \cite{StRaSh} for the first time. 
In other words,
it is expected that the effect of the CEP, if any,
changes only slowly as the collision energy is changed.

Two comments are in order here. First, as we explained,
the universality does not tell us about the sizes of
$\Delta T_{\rm crit}$, $\Delta \mu_{B {\rm crit}}$, $D$,
and so forth. 
If $D$ is larger, the size of the critical region becomes smaller 
and the behavior of the 
isentropic trajectories approaches that in Fig.\ \ref{Fig-nbsc},
which can be seen from 
Eqs.\ (\ref{Eq-size}) and (\ref{Eq-rentro}). 
Namely in the case of larger $D$, the contour lines of $S_c$ 
(for example, $S_c=-1,-0.5,0.5$ and $1$ in Fig.\ \ref{Fig-Scrit}) are 
closer to the $S_c=0$ line, i.e.,
the tangential line to the QCD phase boundary at the CEP.  
Then, in the limit of $D \rightarrow\infty$,
all contour lines shrink to one line, which is identical to
the QCD phase boundary as long as the curvature of the
phase boundary can be ignored. This can be seen from
Eq.\ (\ref{Eq-rentro}).
Therefore, the bag plus excluded volume model can be
considered as an extreme case of 
the equation of state with the CEP. 
Second, the focusing
on the right hand side of the CEP was first discussed
in Ref.\ \cite{StRaSh}. Our numerical result indicates that
the CEP attracts isentropic trajectories not only on
the right hand side (first order side) of the CEP but also
on the left hand side (crossover side) of the CEP. Thus,
if the critical region is large enough, the fine-tuning of
the collision energy is not necessary to hit the vicinity
of the CEP not only on the low energy side \cite{StRaSh}
but also on the high energy side.


\begin{figure}[h]
\begin{minipage}{0.49\linewidth}
\includegraphics[width=1.03\linewidth]{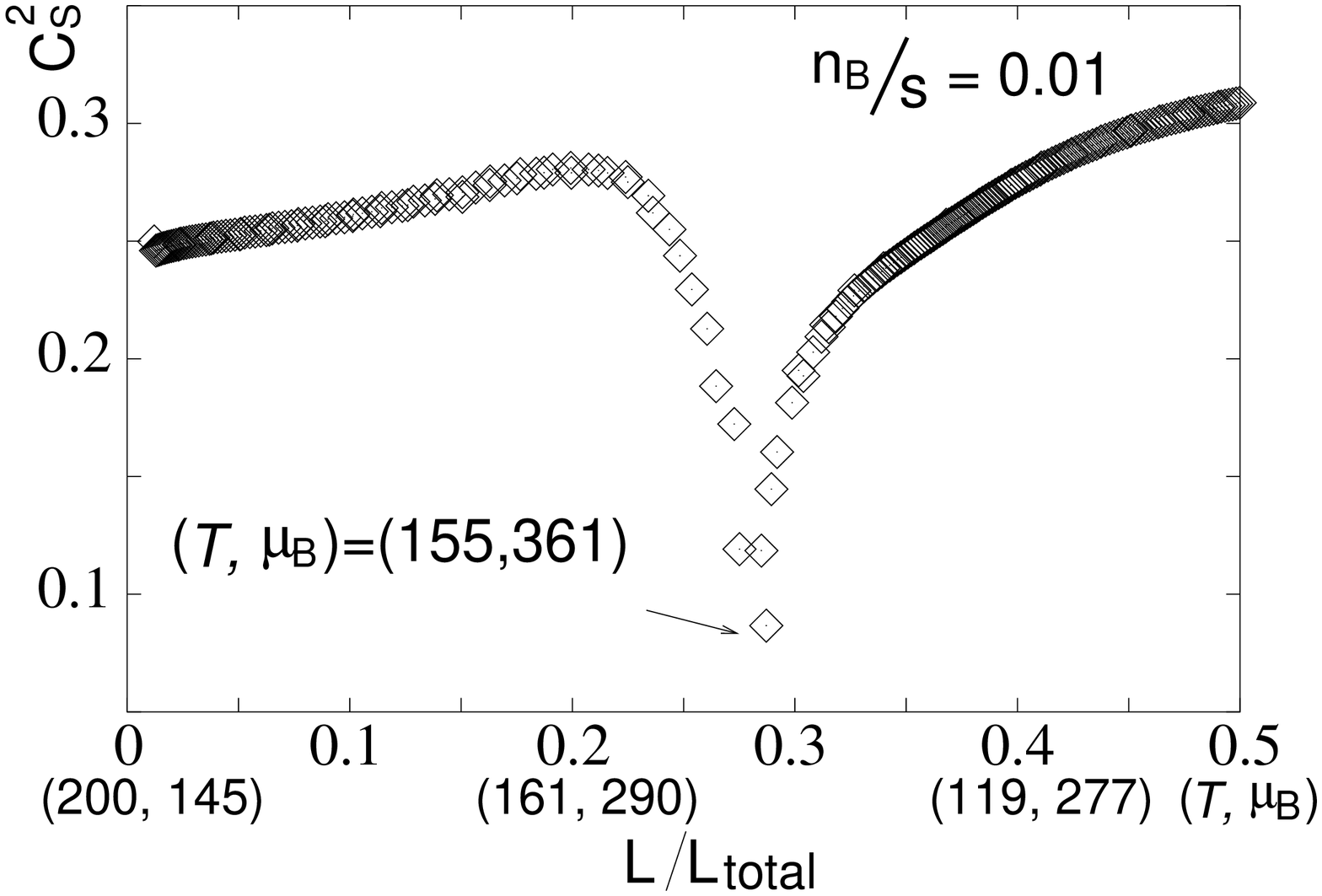}
\end{minipage}
\begin{minipage}{0.49\linewidth}
\vspace{-7mm}
\includegraphics[width=1.03\linewidth]{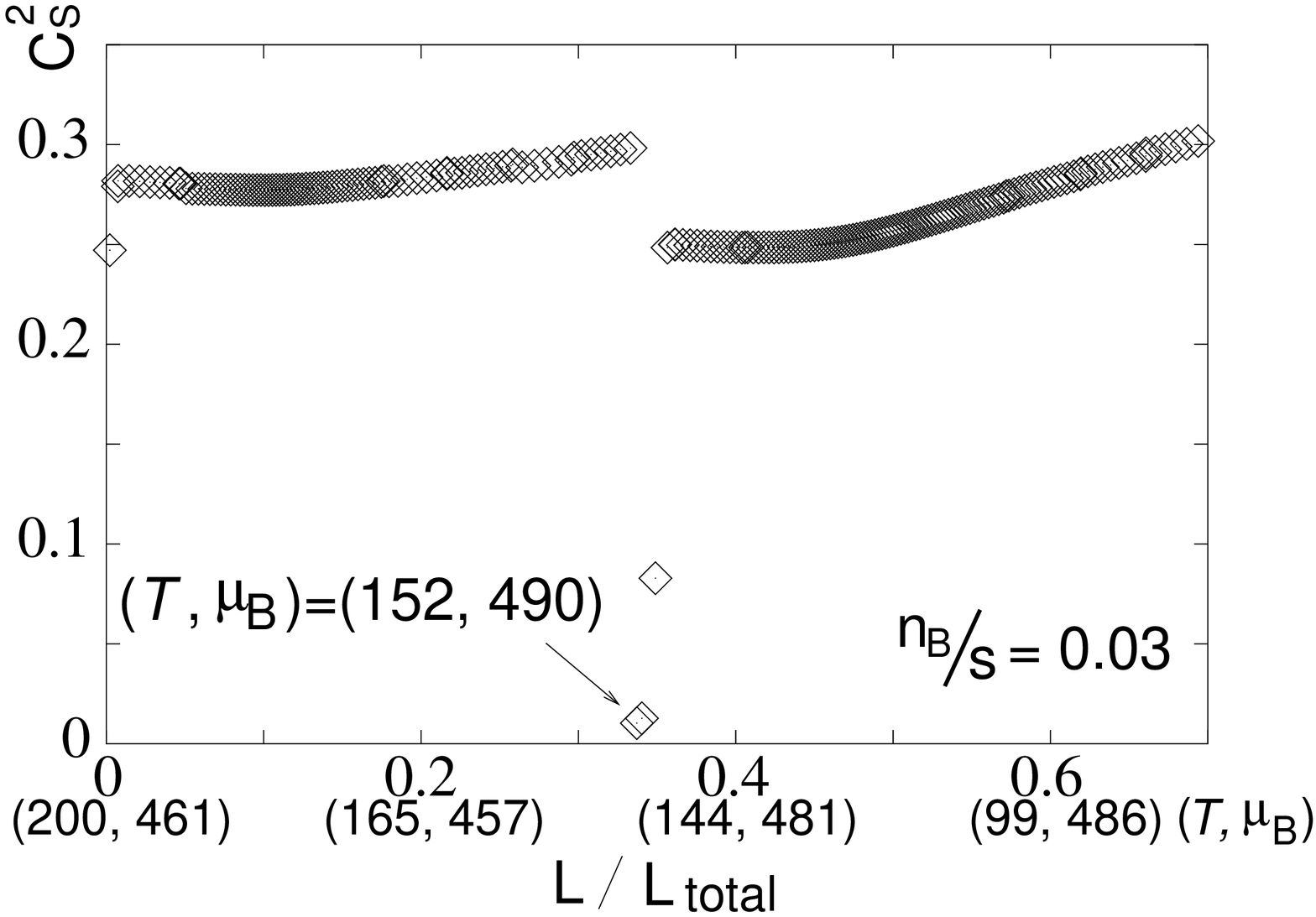}
\end{minipage}
\caption{\label{Fig-cs}
Square of the sound velocity $c_s^2$ as a function of
$L/L_{\rm total}$ along $n_B/s=0.01$
(left) and $n_B/s=0.03$ (right) isentropic trajectories.
The CEP is located at $(T_E, \mu_{BE}) = (154.7 \hspace{2mm}{\rm MeV}, 
~367.8 \hspace{2mm}{\rm MeV})$ in both cases.
The parameters are the same as in Figs.\ 2, 3, and 4 (left).
}
\end{figure}

Next we argue the change of the square of the sound velocity $c^2_s$,
\begin{equation}
c^2_s = \left ( \frac{\partial P}{\partial E} \right )_{n_B/s},
\end{equation}
along the isentropic trajectories,
which gives us the information on how the
equation of state changes in the course of the hydrodynamical evolution.
Figure \ref{Fig-cs} indicates the sound velocity
as a function of $L/L_{\rm total}$,
where $L$ is the path length to a point along the isentropic trajectory
with a given $n_B/s$ from a reference point on the same isentropic
trajectory on the $T$-$\mu_B$ plane,
and $L_{\rm total}$ is
the one to another reference point along the trajectory. 
For example, on the insentropic trajectory $n_B/s=0.01$ in 
Fig.\ \ref{Fig-nbs} (left), $L/L_{\rm total}=0$,
which is the starting point of $n_B/s$ trajectory,
corresponds to the point 
$(T, \mu_B)=(200 \hspace{2mm}{\rm MeV}, 145 \hspace{2mm}{\rm MeV})$ 
and $L/L_{\rm total}=0.29$
corresponds to the point 
$(T, \mu_B)=(155 \hspace{2mm}{\rm MeV}, 361 \hspace{2mm}{\rm MeV})$.
At about
$L/L_{\rm total}=0.25 \sim 0.35 $ the hadronization process 
occurs and  
at about $L/L_{\rm total}=0.35 \sim 0.5$ kinetic freezeout takes place.  
We can see clear difference between the behavior of the sound velocities on
$n_B/s = 0.01$ and 0.03 trajectories.
In the case of $n_B/s=0.01$, the sound velocity changes smoothly along
the trajectory,  which reflects the occurrence of
a smooth phase transition at this $n_B/s$ value, and it takes its minimum
at $L/L_{\rm total}=0.29$.
On the other hand, the sound velocity at $n_B/s=0.03$ changes suddenly
at $L/L_{\rm total}= 0.34$, due to the first order phase transition.
The hydrodynamical expansion along various $n_B/s$ paths
differs in how the effect of the phase transition appears
because of the existence of the CEP.
In a real collision system, $n_B/s$ changes from position to position.
As a result, the time evolution of such a system is described
as a superposition of trajectories with different $n_B/s$'s.
Thus, the system may not expand uniformly and this effect may
appear in physical observables related to expansion such as
the collective flow and Hanbury Brown-Twiss (HBT) radii if the collision
parameters are
appropriate, i.e., isentropic trajectories in the
system pass through and/or near the CEP.

\section{Slowing Out of Equilibrium}

The equilibrium correlation length and fluctuation
become large near the CEP and they diverge at the CEP.
At the same time, 
the typical time scale becomes long
near the CEP, which makes
the system take long time to reach the thermal equilibrium near the CEP.
Therefore, in the time evolution in ultrarelativistic heavy ion collisions,
near the CEP slowing-out-of-equilibrium occurs and
non-equilibrium dynamics has to be taken into account.
We assume that the thermal equilibrium is
established soon after collisions and that the quark-gluon plasma
follows Bjorken's scaling solution \cite{Bj} for hydrodynamical evolution.
The initial temperature and proper time are set to 200 MeV and 1 fm/$c$,
respectively
\footnote{In Bjorken's one-dimensional hydrodynamical evolution,
the non-equilibrium correlation 
length is not sensitive to the choice of the initial conditions.  
Instead, as we discuss later, the reduction of the time scale
due to the transverse expansion is expected to have more effect
on it.}.
Before discussing the correlation length in out-of-equilibrium
time evolution, we calculate the equilibrium 
correlation length near the CEP in the $r$-$h$ plane.
Using Widom's scaling law \cite{BrLeZi},
the equilibrium correlation length $\xi_{\rm eq}$ is given by
\begin{equation}
\xi^2_{\rm eq}(r, M) = a^2 M^{-2 \nu /\beta}
g \left ( \frac{|r|}{|M|^{1/\beta}} \right ), 
\end{equation}
where $\nu=0.63$
is a critical exponent of the
three dimensional Ising model \cite{GuZi}.
$a$ is a constant with the dimension of length and fixed to 1 fm.    
There is not so large ambiguity in the determination of $a$,
since $\xi_{\rm eq}$ is of the order of 1 fm at $T \sim 200$ MeV 
The function $g(x)$ is given by the $\epsilon$ expansion to order 
$\epsilon^2$ \cite{BrLeZi}, 
\begin{eqnarray}
g(x) & = & g_\epsilon (x) \nonumber \\
& = &6^{-2 \nu}z 
\left \{ 
1 - \frac{\epsilon}{36}
[(5+ 6 \ln 3)z-6(1+z)\ln z]
\right. \nonumber \\
& &  + \epsilon^2
     \left [
     \frac{1+2z^2}{72} \ln^2 z +  
     \left (
     \frac{z}{18} \left ( z - \frac{1}{2} \right ) \left( 1-\ln3 \right ) 
     \right .
     \right .        \nonumber \\
& &  - \frac{1}{216} \left. \left. \left
     ( 16 z^2 - \frac{47}{3}z-\frac{56}{3} \right )
     \right ) \ln z  \right .  \nonumber \\
& &  +  \left .\frac{1}{216} \left (
     \frac{101}{6} + \frac{2}{3}I + 6 \ln^2 3 + 4 \ln 3 - 10
     \right ) z^2  \right . \nonumber \\
& &  - \left .\left. \frac{1}{216}\left (
     6 \ln^2 3 + \frac{44}{3} \ln 3 +  \frac{137}{9} + \frac{8}{3}I
     \right )z 
     \right ] \right \}, \nonumber \\
\label{Eq-gepsilon}
\end{eqnarray}
where $z \equiv \frac{2}{1+x/3}$,
$I \equiv \int^1_0\frac{\ln[x(1-x)]}{1-x(1-x)}dx \sim 
-2.344$, and $\epsilon = 4-d$ with $d$ being the dimension of the space, 3.
At large $x$, i.e., around the crossover and CEP,
$g_\epsilon (x)$ cannot be used and the
asymptotic form given in Ref.\ \cite{BeRa}, 
\begin{equation}
g(x)= g_{\rm large}(x) = \left( 
\frac{1}{3+x}
\right) ^{2\nu},
\end{equation}
should be used. 
We smoothly connect $g_\epsilon (x)$ and $g_{\rm large} (x)$ around $x=5$, 
i.e., in most part of the critical region $g(x)$ is given by
Eq.\  (\ref{Eq-gepsilon}).  

\begin{figure}[h]
\begin{minipage}{0.49\linewidth}
\includegraphics[width=1.0\linewidth]{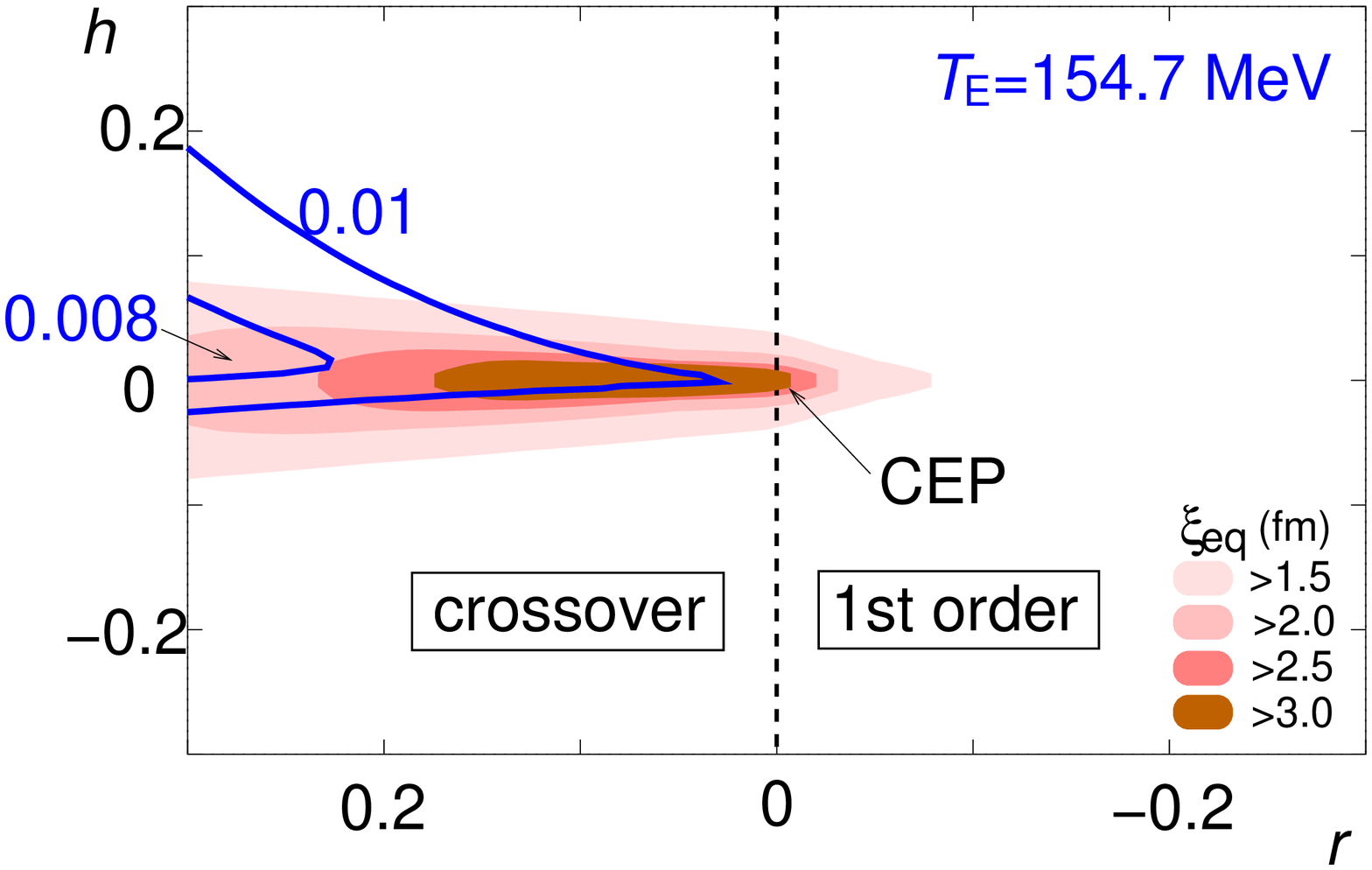}
\end{minipage}
\begin{minipage}{0.49\linewidth}
\includegraphics[width=0.95\linewidth]{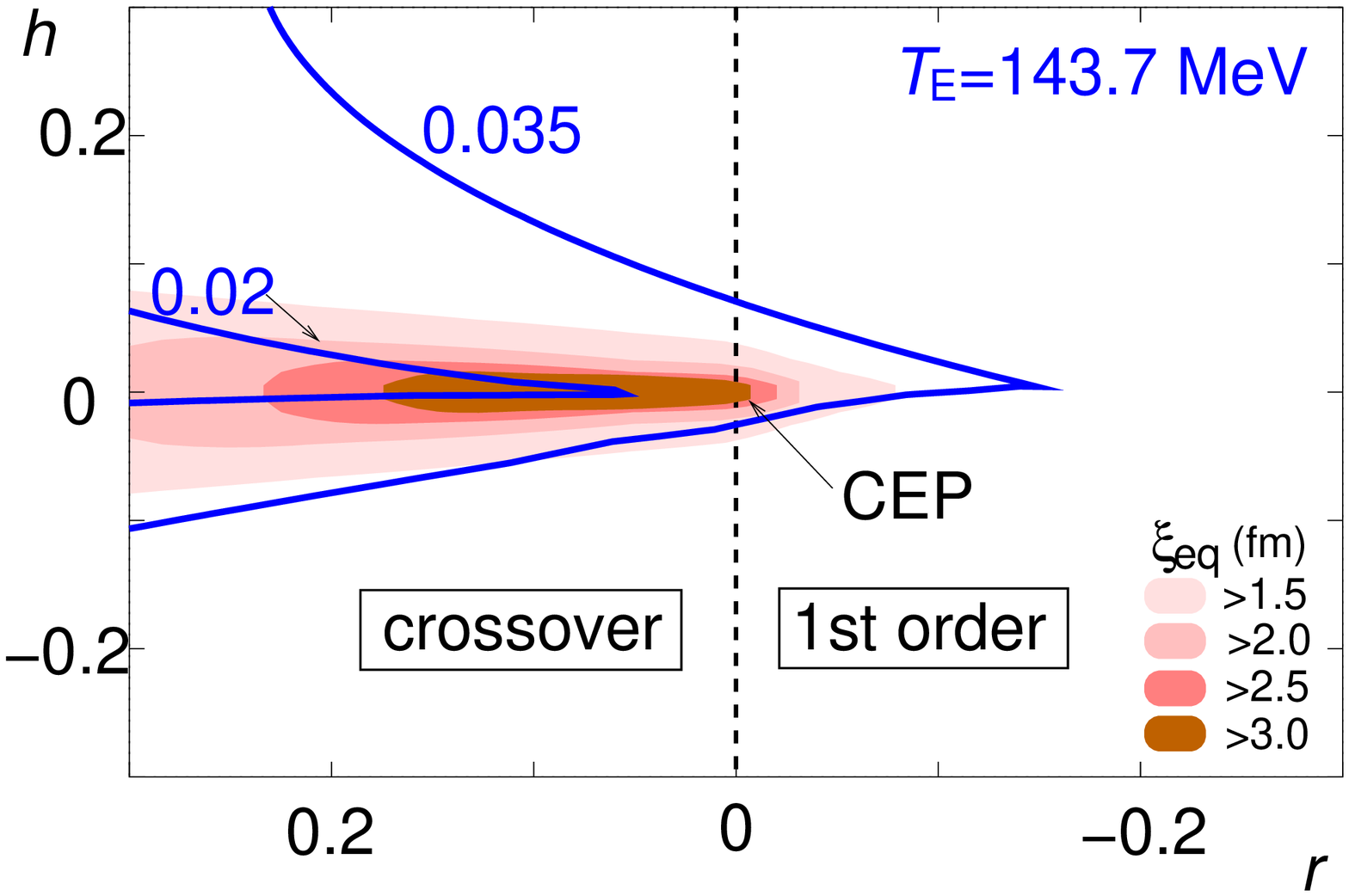}
\end{minipage}
\caption{\label{Fig-coreq}
Correlation length in equilibrium and the $n_B/s$ contour lines in 
the $r$-$h$ plane.
The solid lines are trajectories with fixed $n_B/s$ values
0.008 and 0.01 in the case of $(T_E, \mu_{BE}) = (154.7 \hspace{2mm}{\rm MeV}, 
~367.8 \hspace{2mm}{\rm MeV})$ (left) 
and 0.02 and 0.035 in the case of $(T_E, \mu_{BE}) = 
(143.7 \hspace{2mm}{\rm MeV}, ~652.0 \hspace{2mm}{\rm MeV})$ (right).  
}
\end{figure}

Figure \ref{Fig-coreq} shows the correlation length in equilibrium 
and the isentropic trajectories at some $n_B/s$'s in the $r$-$h$ plane. 
In Fig.\ \ref{Fig-coreq} the CEP is located at the origin, and
the first order phase transition occurs at 
$r < 0$ and the crossover phase transition takes place at $r > 0$.
The correlation length in equilibrium $\xi_{\rm eq}$ is divergent at the 
origin and the region where $\xi_{\rm eq}$ is large spreads out in the 
smooth phase transition region, $r > 0$.       
In Fig.\ \ref{Fig-coreq} solid lines stand for the
isentropic trajectories in the case of
$(T_E, \mu_{BE})=(154.7 \hspace{2mm}{\rm MeV}, 
~367.8 \hspace{2mm}{\rm MeV})$ (left) and    
($T_E, \mu_{BE})=(143.7 \hspace{2mm}{\rm MeV}, 
~652.0 \hspace{2mm}{\rm MeV})$ (right).    
Note that this $r$-$h$ plane is actually placed in the $T$-$\mu_B$
plane with the tilt given by the slope of the tangential line of
the first order phase transition line
at the CEP as shown in Fig.\ \ref{Fig-map}. The value of
$n_B/s$ is determined not only by $r$ and $h$ but also by 
the non-critical component of the equation of state.
This is the reason why the behavior of the trajectories
relative to the CEP changes 
according to the location of the CEP in the $T$-$\mu_B$ plane. 
Thus, the strength of focusing effect of the CEP 
is determined also by the non-singular part.         
On the other hand, we 
use the universality hypothesis to calculate $\xi_{\rm eq}$ and,
as a result,
it is a function of only $r$ and $h$.
The trajectories are attracted to the CEP, 
which makes the length of each trajectory in the critical region,
where $\xi_{\rm eq}$ is large, long. 
This behavior is very different from the assumption
in usual schematic analyses, for example, Fig.\ 1 in
Ref.\ \cite{BeRa} by Berdnikov and Rajagopal.
The maximum value of $\xi_{\rm eq}$ depends on the value of $n_B/s$.

Next we calculate the non-equilibrium correlation length $\xi(\tau)$
as a function of the proper time $\tau$ by using the rate equation
given in Ref.\ \cite{BeRa}, 
\begin{equation}
\frac{d}{d\tau}m_\sigma(\tau)=-\Gamma[m_\sigma(\tau)] 
\left (
m_\sigma(\tau) - \frac{1}{\xi_{\rm eq}(\tau)}
 \right ), 
\end{equation}
where $1/m_\sigma(\tau)=\xi(\tau)$ and $\Gamma[m_\sigma(\tau)]$
is the parameter which represents the rate of slowing-out-of-equilibrium.
From the theory of dynamical critical phenomena,
$\Gamma[m_\sigma (\tau )]$ is given as
\begin{equation}\label{xi-evo}
\Gamma[m_\sigma (\tau )]=\frac{A}{\xi_0}(m_\sigma(\tau) \xi_0)^z,
\end{equation}
where we take $\xi_0 = \xi_{\rm eq}(T=200~{\rm MeV})$, and
$z$ is a universal exponent of Model H in \cite{HoHa77}, i.e.,
$z \sim 3$ \cite{BeRa,SoSt04}.
$A$ is a dimensionless non-universal parameter and we use $A=1$
as in Ref.\ \cite{BeRa}, though
there is considerable uncertainty in
the parameter $A$. However, fortunately, 
the relation between $\xi_{\rm eq}$ and $\xi$ does not depend on
the parameter choice of $A$ so much at the kinetic freezeout point.
The non-equilibrium correlation length approaches the equilibrium correlation
length as $A$ increases. At $A = 100$,
$\xi$ is almost equal to
$\xi_{\rm eq}$. On the other hand, at small $A$ the effect of
slowing-out-of-equilibrium appears strongly.
At the same time, however, the maximum value of $\xi$ becomes small
as $A$ decreases.
As a result, the absolute value of the difference between $\xi_{\rm eq}$ and
$\xi$ at the kinetic freezeout temperature remains almost unchanged
for a wide range of $A$.
Indeed, we have checked that the following discussion about non-equilibrium
correlation length holds for
$0.1 \lesssim A  \lesssim 10$.

\begin{figure}
\includegraphics[width=0.95\linewidth]{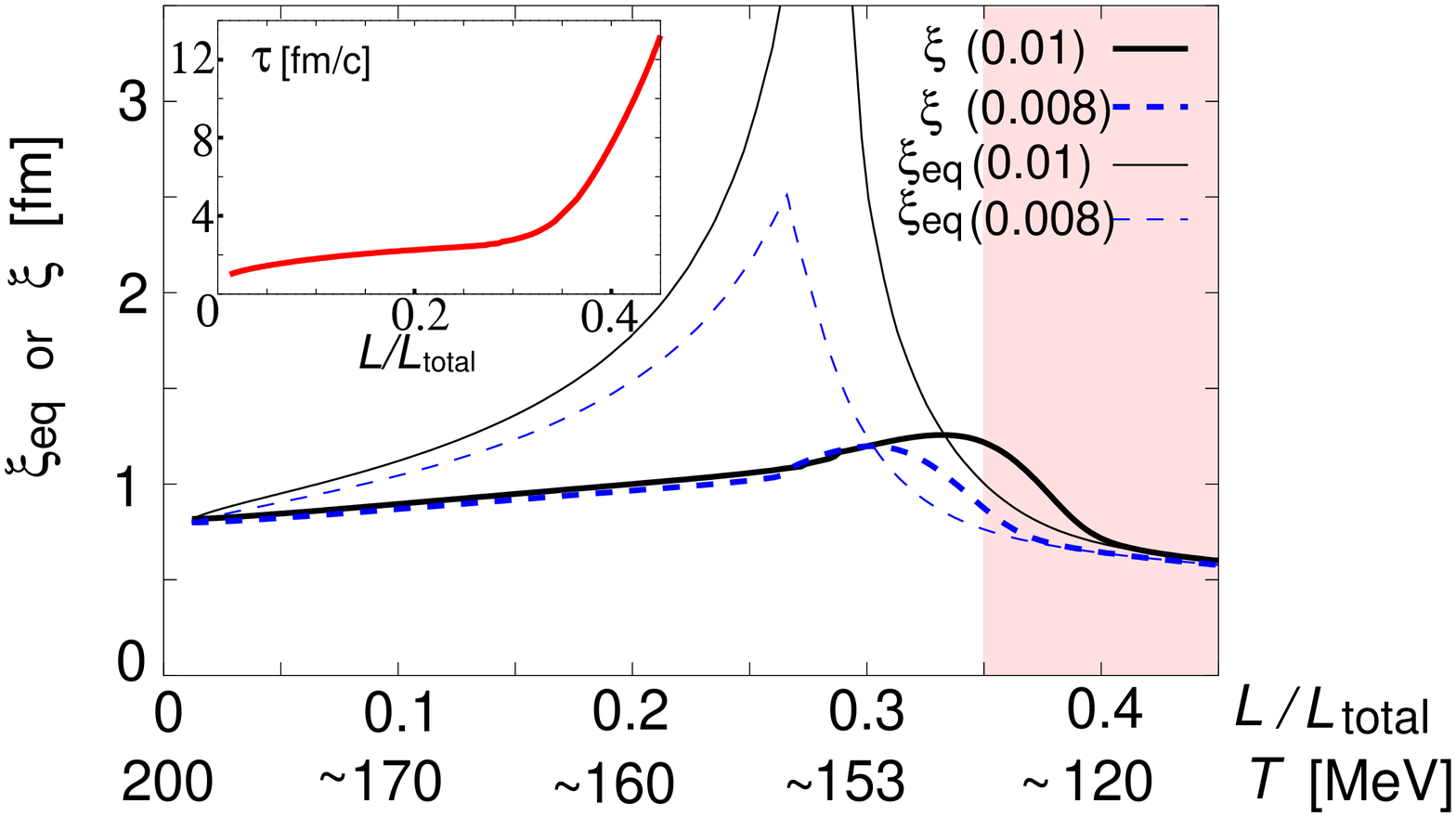}
\caption{\label{Fig-cor} 
Equilibrium correlation length $\xi_{\rm eq}$ (thin lines) and 
non-equilibrium correlation length $\xi$ (thick lines)
on the isentropic trajectories
with $n_B/s=0.008$ and $0.01$, together with $\tau$ as functions 
of $L/L_{\rm total}$ (inlet). $z=3$ was used in the calculation.
}
\end{figure}

\begin{figure}
\includegraphics[width=0.95\linewidth]{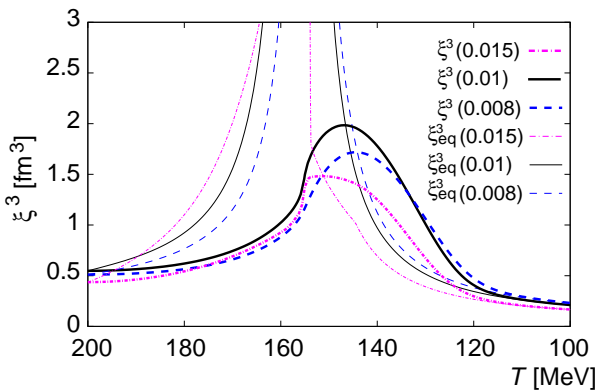}
\caption{\label{Fig-corcube} Cube of the  equilibrium correlation length (thin lines) 
as a function of the temperature and the cube of the    
non-equilibrium correlation length (thick lines)
on the isentropic trajectories
with $n_B/s=0.008$, $0.01$, and $0.015$.
The parameters are the same as for Fig.\ \ref{Fig-cor}.
The isentropic lines with $n_B/s =0.008$, $0.01$, and $0.015$
pass left of, almost through, and right of the CEP, respectively.}
\end{figure}

Figure \ref{Fig-cor} shows the correlation length as a function of 
$L/L_{\rm total}$.
The dashed and solid lines stand for the correlation lengths 
in equilibrium at $n_B/s=0.008$ and $n_B/s=0.01$,
respectively. 
The parameters are the same as for Fig.\ \ref{Fig-nbs} (left).
The thin and thick lines are $\xi_{\rm eq}$ and $\xi$,
respectively. The maximum value of $\xi_{\rm eq}$ along the former
trajectory is larger than that along the latter, because the 
former approaches the CEP more closely than the latter.
The non-equilibrium correlation length $\xi$ is smaller than
$\xi_{\rm eq}$ at the beginning.
Then, $\xi$ becomes larger than $\xi_{\rm eq}$ later.
These are both due to the critical slowing down around the CEP,
as pointed out in Ref.\ \cite{BeRa}. However, the difference becomes
small by the time the system gets to the
kinetic freezeout point. If the transverse expansion is taken
into account, the time scale in the hadron phase becomes much shorter,
but $\xi_{\rm eq}$ is already small in the hadron phase and
the difference is expected to remain small.
Figure \ref{Fig-corcube} shows the cube of the correlation length
as a function of the temperature for $n_B/s=0.008$, $0.01$, and $0.015$.
The parameters are the same as for Fig.\ \ref{Fig-cor}.
The fluctuation of the sigma field scales approximately
as $\propto \xi^3$. The singular part of the fluctuation of intensive
thermal quantities of pions near the CEP is proportional to
$\xi^2$ \cite{StRaSh}. The general relation between $\xi$ and
observed fluctuations still needs to be clarified. These relations,
however, suggest that there is room
for the observation of some enhancement in fluctuations when the
system passes the vicinity of the CEP even if the enhancement of
$\xi$ is small.

 
\section{Consequences of the Critical End Point in Experiments}

Many experimental analyses have been carried out in search of
the evidence of the CEP in the QCD phase diagram. Since the correlation
length and fluctuation diverge at the CEP in thermal equilibrium,
it has been expected that 
some enhancement of the fluctuation, for instance,
is observed if the collision energy of nuclei is
properly adjusted so that the system goes right through the CEP.
Generally, the higher the collision energy is,
the smaller chemical potential region is explored in the
$T$-$\mu_B$ plane. Thus, it has also been expected that
the observables related to the critical behavior around the
CEP such as fluctuations show non-monotonic behavior as a function
of the collision energy. We discuss the observability of
such behavior in ultrarelativistic heavy ion collisions on the
basis of our findings in the previous sections.

\begin{figure}
\includegraphics*[width=0.8\linewidth,angle=0,clip=true]{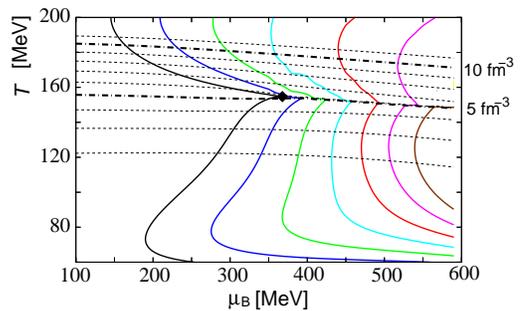}
\caption{\label{Fig-s-cont-CEP} 
Isentropic trajectories and entropy density contour lines in the case
with the CEP. The parameters are the same as for Fig.\ \ref{Fig-EOS}
and the left figure of Fig.\ \ref{Fig-nbs}. The solid lines are
isentropic trajectories
($n_B/s =$ 0.01, 0.015, 0.02, 0.025, 0.03, 0.035, and 0.04 from
left to right) as in Fig.\ \ref{Fig-nbs}. The dashed lines are entropy
density contour lines, which are shown at every 1 fm$^{-3}$.
}
\end{figure}

It has been naively expected that the collision energy needs to be
carefully adjusted so that the system goes right through the CEP
and its existence can be confirmed. In section II, we have shown
that the CEP acts as an attractor of the isentropic trajectories.
Thus, if the size of the critical domain
is large enough, as stressed in Ref.\ \cite{StRaSh}
for the right hand side of the CEP, 
it is not necessary to
fine-tune the collision energy to make the system approach
the CEP closely enough. However, such physical quantities
that show the critical behavior near the CEP are hadronic
observables, and are subject to final state interactions
in the hadron phase. It has been often argued that if the system
passes near the CEP, kinetic freezeout takes place near
the CEP \cite{StRaSh}. It is based on the expectation that, in such a
case, the system stays long near the CEP, where the phase transition
is second order, and that when the system starts to leave away
from the CEP, the entropy density or particle density is already
small enough so that kinetic freezeout takes place. To the contrary,
Fig.\ \ref{Fig-EOS} does not show a sharp drop of the entropy density
near the CEP. To see this more clearly, we show the contour plot of
the entropy density in Fig.\ \ref{Fig-s-cont-CEP} for the same
parameters as in Fig.\ \ref{Fig-EOS}. The entropy density gives
a semi-quantitative measure of the whereabouts of
kinetic freezeout. Alternatively, we could use the energy
density as a measure as well. Figure \ref{Fig-s-cont-CEP} shows that
the contour lines of the entropy density are not focused near
the CEP in contrast to the isentropic trajectories. Thus, it is not
likely that kinetic freezeout takes place near the CEP.
The reason can be traced back to the fact that at the CEP the
entropy density is continuous
and bound, although its derivative along the $h$ axis diverges.
If the system does not freeze out near the CEP,
the dilution of the critical behaviors in the hadron phase
needs to be carefully considered.

Empirically it has been known that chemical freezeout takes
place just below the theoretically expected phase transition line.
Since the isentropic trajectories are focused around the CEP,
it would be tempting to expect that the chemical freezeout points
are focused as the collision energy is varied. However, so far,
only the free resonance gas model has been used in the analysis of
the chemical freezeout point. If the critical region is so large that the
focusing is realized in a large region,
the effect of the interactions on hadrons
can by no means be ignored. Then, the hadrons at chemical freezeout
are at most quasi-particles which are quantum-mechanically different
from the hadrons in the vacuum \cite{AsCsGy} and it is necessary
to consider the time evolution of the quasi-particle states to relate
the chemical freezeout points and the observed particle number ratios.
This is beyond the scope of this paper and we would
like to leave this as a future problem.

\begin{figure}
\includegraphics*[width=0.8\linewidth,angle=0,clip=true]{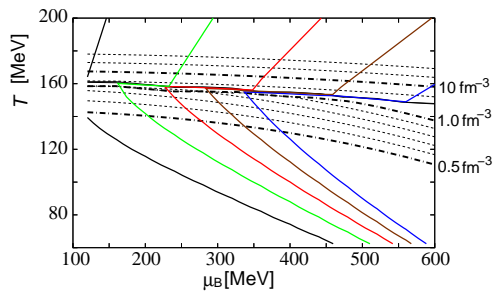}
\caption{\label{Fig-s-cont-excl} 
Isentropic trajectories and entropy density contour lines in the case
without the CEP. The parameters are the same as in Fig.\ \ref{Fig-nbsc}.
The solid lines are isentropic trajectories
($n_B/s =$ 0.01, 0.02, 0.03, 0.04, and 0.05 
from left to right) as in Fig.\ \ref{Fig-nbsc}.
The dashed lines are the entropy density contour lines.
The contour lines are shown at every 0.1 (1)
fm$^{-3}$ in the hadron (QGP) phases because the increase rate of 
the entropy density in the QGP phase is much larger than that in
the hadron phase due to the strong first order phase transition.  
}
\end{figure}

The entropy density 
in Fig.\ \ref{Fig-EOS} shows the following feature.
On the left hand side of the CEP, the entropy density changes smoothly,
reflecting the fact that the phase transition here is crossover.
The closer to the $T$ axis, the less rapid the change is.
On the other hand, on the right hand side of the CEP, the phase transition
is of first order, and the jump in the entropy density increases
as the chemical potential increases. As a result, i) the separation
of the contour lines of the entropy density and energy density gets
larger as the distance from the CEP becomes larger on the left hand
side of the CEP,
ii) the contour lines below the crossover are flatter
than those in the hadron phase in the case with a first order
phase transition (See Figs.\ \ref{Fig-EOS} and \ref{Fig-s-cont-CEP}).
This feature is not due to the universality. 
Instead, it is a general consequence of the existence of the CEP.
In fact, as shown in Fig.\ \ref{Fig-s-cont-excl}, this feature
is not observed in the case of the bag plus excluded volume
model. Using this feature of the entropy density, 
we discuss the behavior of kinetic freezeout as a function
of the collision energy. Kinetic freezeout takes place when the mean
(elastic) collision rate and the expansion rate of the system become
comparable. The expansion rate in central collisions at RHIC is known to
be considerably larger than that at SPS \cite{RHIC-freezeout}.
The elastic collision rate can be estimated from the 
elastic cross sections for pions and baryons and their densities.
Due to the chiral symmetry and the effect
of the $\Delta$, the elastic $\pi$-$N$ cross section is considerably
larger than that of $\pi$-$\pi$.
On the other hand, in Pb+Pb collisions at 158 AGeV at SPS,
$ (p +\bar{p})/(\pi^++\pi^-) \sim 0.09$
\cite{NA4904,SoHeSoXe98}
and it is $ \sim 0.09$ at RHIC
\cite{STARQM02}. 
Thus, at the same particle number density,
the collision rate at SPS is almost the same as that at RHIC.
From these considerations, it would be natural if the
kinetic freezeout temperature in central collisions at RHIC is
noticeably higher than that at SPS, while the kinetic freezeout temperature
at RHIC is actually slightly lower than that at SPS
\cite{RHIC-freezeout}. 
Two comments are in order here. 
First, the kinetic freezeout we are considering
here is that of pions and kaons. The analysis in Ref.\ \cite{RHIC-freezeout}
does not take into account the effect of resonance decays. 
When resonance decays are taken into account, the kinetic freezeout
temperature becomes even lower \cite{Kiyo}. 
Second, more precisely speaking,
kinetic freezeout takes place when the mean collision rate
and the product of the expansion rate and the typical scale of the system
become comparable. However, according to the results of the 
HBT interferometry, the typical size at RHIC is not increased as
the expansion rate compared to SPS.

This can be naturally understood
if the separation of the entropy density ($\sim$ particle density)
or energy density contour lines is larger at kinetic freezeout at RHIC
than at SPS. This implies that the CEP exists in the QCD phase diagram and,
furthermore, that the RHIC isentropic trajectory passes on the left hand side
of the CEP. To confirm this tendency, it is necessary to obtain
the kinetic freezeout temperatures at several collision energies
between the SPS and current RHIC energies.

In the previous section, we have shown that the difference between the
thermal correlation length $\xi_{\rm eq}$ and the actual correlation length
$\xi$ is small at
kinetic freezeout as well as $\xi$ itself. 
As explained in Ref.\ \cite{BeRa}, precisely speaking,
Eq.\ (\ref{xi-evo}) is valid only near the CEP. Nevertheless, the
physical reason why $\xi_{\rm eq}$ and $\xi$ are so close to each other
at kinetic freezeout, is general and holds also in this case;
final state interactions tend to wash out non-equilibrium effects.
Since the kinetic freezeout points at the Super Proton 
Synchrotron (SPS) at CERN and the Relativistic Heavy Ion Collider
(RHIC) at Brookhaven National Laboratory are far below
the phase transition line, the effect of final state interactions
is expected to dominate. 

Similar arguments hold also for fluctuations
except for the cube effect we argued in the previous section.
The event-by-event fluctuations of the mean transverse momentum in
Pb + Au collisions at 40, 80, and 158 AGeV/$c$ were measured by
the CERES Collaboration \cite{CERES03}.
They are slightly smaller than those at RHIC, and unusually large
fluctuation or non-monotonic behavior which may suggest the existence of the
CEP has not been observed.
However, in order to understand experimental observations,
we have to take account of the CEP character as
an attractor of isentropic trajectories in the $T$-$\mu_B$ plane.
Due to the focusing effect of the CEP, the isentropic trajectories
of various initial collision energies are gathered to the CEP,
if the critical region around the CEP is large enough.
As a result, when each isentropic trajectory passes near the CEP,
similar correlation lengths, fluctuations, and so on are induced,
and they do not show strong non-monotonic behavior
as a function of the collision energy.
In other words, the absence of the non-monotonic behavior,
which is shown by the CERES Collaboration, does not necessarily
imply the non-existence of the CEP in the region probed by SPS and RHIC.
More precise measurement and analyses which take account of the
focusing and slowing-out-of-equilibrium may reveal the existence
of the CEP.

Recently, the fluctuation of $(K^+ + K^-)/(\pi^+ + \pi^- )$ and
$(p + \bar{p} )/(\pi^+ + \pi^- )$
was measured and it was found that the collision energy
dependence of the fluctuation of $(K^+ + K^-)/(\pi^+ + \pi^- )$
is not described by a cascade model \cite{NA4904}.
While this result is interesting, we have to be careful in
associating this result with the critical phenomena around the CEP.
Since neither the numerator nor denominator of
$(K^+ + K^-)/(\pi^+ + \pi^- )$ is a conserved quantity in
the strong interaction, substantial modification is expected for
both in the hadron phase as we have demonstrated for the correlation
length. Furthermore, as we discussed for the particle ratios,
quasi-particle states near the CEP are quantum-mechanically different
from the ones in the vacuum, i.e., observed pions, kaons, protons, and so on.
To relate the fluctuation of the particle ratios to the critical phenomena,
it is necessary to know not only the time evolution of the quasi-particle
states but also the relation between the quasi-particle quantum states
in medium and the particle states in the vacuum \cite{AsCsGy}.

\begin{figure}
\includegraphics*[width=0.9\linewidth]{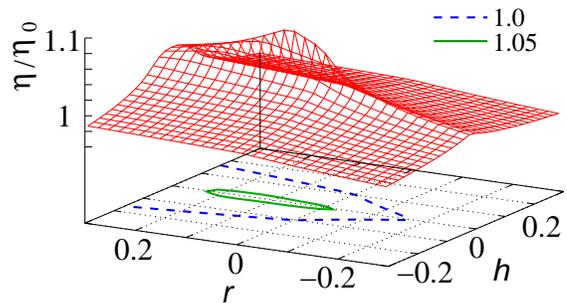}
\caption{\label{Fig-vis} 
Shear viscosity $\eta/\eta_0$ near the CEP. For the parameters used in the
calculation, see the text.
}
\end{figure}

Up to now, we have been considering the isentropic trajectories
in the $T$-$\mu_B$ plane. Actually, the viscosity is known to
diverge at the CEP and the entropy is generated if the system passes
near the CEP. However, the singularity in the viscosity
is known to be very weak \cite{SoSt04},
\begin{equation}
\frac{\eta}{\eta_0} = \left (\frac{\xi_{\rm eq}}{\xi_{\rm eq,0}} \right )^
{(\frac{1}{19} \epsilon + O(\epsilon^2))},
\end{equation}
where $\eta$ is the shear viscosity, $\eta_{\rm 0}$ is
the shear viscosity at the point where the equilibrium
correlation length is $\xi_{\rm eq,0}$, and 
$\epsilon$ is $4-d$ as before.
In Fig.\ \ref{Fig-vis},
$\eta/\eta_0$ in the $r$-$h$ plane is shown.
The small correction to the critical exponent
$1/19$ is set to 0 for simplicity. Because of the small critical exponent,
the diverging feature of the viscosity is hardly seen. The effect
of the entropy production around the CEP is thus expected to be small, and
it will not show easily-recognizable non-monotonic changes in
observables as the collision energy is changed.

\section{Summary}
In this paper,
we discussed the hydrodynamical expansion near the QCD critical end point 
and its consequences in experimental observables.      
First we constructed realistic equations of state with the CEP on 
the basis of the universality hypothesis and discussed the hydrodynamical 
expansion near the CEP. 
The behavior of isentropic trajectories near the QCD critical end point 
is clearly different from that in the bag plus excluded volume model, which 
is usually used in hydrodynamical models. 
We found that the CEP acts
as an attractor of isentropic trajectories, $n_B/s = {\rm const.}$,
in the $T$-$\mu_B$ plane, 
not only on the right hand side of the CEP but also on the 
left hand side of the CEP.
Because of the focusing of the isentropic trajectories,
the path of the system in the course of the time evolution  
in the $T$-$\mu_B$ plane  
is remarkably different from that in the usual bag plus
excluded volume model.
This will be reflected in the final state observables that are
sensitive to the equation of state, such as the anisotropic flows
(directed, elliptic, and higher order Fourier coefficients)
\cite{Rischke95,Kolb03} and HBT radii.
For this purpose, 
full 3-D hydrodynamical calculations \cite{NoHoMu,Hi02} with
realistic equations of state are indispensable \cite{Nonaka04}.

Next we argued the critical slowing down near the QCD critical end point.
We compared the equilibrium and non-equilibrium correlation lengths
along two isentropic trajectories which pass near the CEP.
We found that the difference between
them at kinetic freezeout is very small.

Furthermore, we considered the experimental observability
of the consequences of the CEP.
Fluctuations and viscosities diverge at the CEP, but fluctuation which 
is induced by the CEP will
fade by kinetic freezeout for the same reason as for
the correlation length. This explains the non-observation of sharp
non-monotonic behaviors in fluctuations.   
However, the singular part of the fluctuations is approximately
proportional to the cube of the correlation length and
there is a possibility that
more precise measurements in the future may reveal
the existence of the CEP. The singularity in the viscosity is
too weak to affect observables. 
Concerning the chemical freezeout process, we need to  
consider the effect of the interactions on hadrons for detailed 
discussion on particle number ratios, when it takes place near the CEP.
We found that, at least now, the most promising way is to trace  
the behavior of the collision energy dependence of the kinetic freezeout
temperature. The fact that the kinetic freezeout temperature
at RHIC is lower than that at SPS suggests the existence of
the CEP in the $T$-$\mu_B$ plane and that the RHIC path goes left of
the CEP. 

Although our discussion in this paper does not depend on
the parameter choice in Eqs.\ (\ref{Eq-size}) and (\ref{Eq-rentro})
qualitatively, our assumption that the critical region near the CEP
spreads out a large area may look too optimistic.
However, according to recent lattice calculations at zero baryon chemical
potential, the $\bar{\psi}\psi$ susceptibility shows a very broad peak
as a function of the temperature. Furthermore, the width tends to even
increase as the lattice spacing is decreased \cite{MILC}.
This clearly indicates
that the free-gas description cannot be used at
$100 \lesssim T \lesssim 250$ MeV at zero baryon chemical potential.

\vspace*{0.5cm}
\begin{acknowledgments}
We thank B. M\"uller for illuminating discussions and E. Shuryak,
M. Stephanov, and K. Rajagopal
for comments. Also, we are grateful to O. Miyamura
and S. Muroya for encouragement.
C.N. was in part supported by DOE grants
DE-FG02-96ER40945 and DE-FG02-03ER41239,
and M.A. was in part supported by
Grant-in-Aid by the Japanese Ministry of Education
No. 14540255.
\end{acknowledgments}

\appendix

\section{stability condition}
In an isolated system, the entropy $S$ takes the maximum value in the 
stable equilibrium state and the following relations \cite{LaLi,GlPr},
\begin{eqnarray}
\delta S &= &0, \\ 
\label{Eq-ds}
\delta ^2 S & <  & 0, 
\label{Eq-d2s0}
\end{eqnarray}
are satisfied.
Here the independent variables are the energy     
$E$, volume $V$, and particle number $N$.     
Suppose that the total system whose volume is $V$ is composed of two parts
with the volumes $V_1$ and $V_2$. Suppose that in $V_1$ ($V_2$)
the variation of the energy, volume, and particle number are  
$\delta E_1$ ($\delta E_2$), $\delta V_1$ ($\delta V_2$), and $\delta N_1$ 
($\delta N_2$), respectively. These variations satisfy     
\begin{eqnarray}
\delta E_1 + \delta E_2 & = & 0,  \nonumber  \\
\delta V_1 + \delta V_2 & = & 0,  \nonumber \\
\delta N_1 + \delta N_2 & = & 0. 
\label{Eq-devn}
\end{eqnarray}
By expanding $\delta S$ with regard to $E$, $V$, and $N$
in Eq.\ (\ref{Eq-ds}) and substituting
$\left ( \frac{\partial S}{\partial E} \right )_{V,N}=\frac{1}{T}$,
$\left ( \frac{\partial S}{\partial V} \right )_{E,N}=\frac{P}{T}$,
$\left ( \frac{\partial S}{\partial N} \right )_{E,V}=-\frac{\mu}{T} $,
and Eq.\ (\ref{Eq-devn}) into Eq.\ (\ref{Eq-ds}), we obtain  
$T_1 =  T_2$,  $ P_1 =  P_2$, and $\mu_1  = \mu_2$,
i.e., both parts of the system have the same temperature, pressure, and 
chemical potential in equilibrium. 

Next we evaluate the second variation $\delta^2 S$ in Eq.\ (\ref{Eq-d2s0}),
\begin{widetext}
\begin{eqnarray}
\delta ^2 S & = &  \left ( \frac{\partial^2 S_1}{\partial E^2_1} 
\right )_{V_1, N_1} \delta^2 E_1 +
\left ( \frac{\partial^2 S_1}{\partial V^2_1} 
\right )_{E_1, N_1} \delta^2 V_1 + 
\left ( \frac{\partial^2 S_1}{\partial N^2_1} 
\right )_{E_1, V_1} \delta^2 N_1   \nonumber \\ 
& & + 
2 \left ( \frac{\delta^2 S_1}{\delta E_1 \delta V_1} 
\right )_{N_1} \delta E_1 \cdot \delta V_1 + 
2 \left ( \frac{\delta^2 S_1}{\delta V_1 \delta N_1} 
\right )_{E_1} \delta N_1 \cdot \delta V_1 +
2 \left ( \frac{\delta^2 S_1}{\delta E_1 \delta N_1} 
\right )_{V_1} \delta E_1 \cdot \delta N_1 \nonumber \\
& & + (1 \rightarrow 2).
\label{Eq-d2s}
\end{eqnarray}
\end{widetext}
By calculating $\delta^2 S$
using $T\delta S = \delta E + P\delta V - \mu \delta N$ and comparing it
with (\ref{Eq-d2s}), we obtain
\begin{equation}
T\delta^2 S = -\delta T_1 \delta S_1 + \delta P_1 \delta V_1 - 
\delta \mu_1 \delta N_1 + (1 \rightarrow 2).
\label{Eq-Td2s}
\end{equation}
By definition, 
\begin{widetext}
\begin{eqnarray}
\delta S_1 & =& \left ( \frac{\partial S_1}{\partial T_1} \right )_{V_1, N_1}
\delta T_1 
+ \left ( \frac{\partial S_1}{\partial V_1} \right )_{T_1, N_1} \delta V_1 + 
\left ( \frac{\partial S_1}{\partial N_1} \right )_{T_1, V_1} \delta N_1,  
\nonumber \\
\delta P_1 & = &   
\left ( \frac{\partial P_1}{\partial T_1} \right )_{V_1, N_1}
\delta T_1 
+ \left ( \frac{\partial P_1}{\partial V_1} \right )_{T_1, N_1} \delta V_1 + 
\left ( \frac{\partial P_1}{\partial N_1} \right )_{T_1, V_1} \delta N_1,   
\nonumber \\ 
\delta \mu_1 & = & 
\left ( \frac{\partial \mu_1}{\partial T_1} \right )_{V_1, N_1}
\delta T_1 
+ \left ( \frac{\partial \mu_1}{\partial V_1} \right )_{T_1, N_1} \delta V_1 + 
\left ( \frac{\partial \mu_1}{\partial N_1} \right )_{T_1, V_1} \delta N_1 .
\label{Eq-dpm}
\end{eqnarray}
\end{widetext}
Substituting Eq.\ (\ref{Eq-dpm}) and the following relations,  
$\left ( \frac{\partial S}{\partial T} \right )_{V,N} = \frac{1}{T} C_v$, 
$\left ( \frac{\partial S}{\partial V} \right )_{T,N} 
= \left ( \frac{\partial P}{\partial T} \right )_{V,N}$,
$\left ( \frac{\partial \mu}{\partial V} \right )_{T,N}
=- \left ( \frac{\partial P}{\partial N} \right )_{T,V}$, 
and $ \left ( \frac{\partial \mu}{\partial T} \right )_{V,N}
=- \left ( \frac{\partial S}{\partial N} \right )_{T,V}$ 
into Eq.\ (\ref{Eq-Td2s}),
and using Eq.\ (\ref{Eq-devn}), we obtain the quadratic form    
\begin{widetext}
\begin{equation}
\label{Eq-d2s2}
T \delta ^2 S \ 
 =  - \frac{1}{T}C_v (\delta T) ^2
+  \left ( \frac{\partial P}{\partial V} \right )_{T,N} 
\left [\delta V + 
\frac{ \left ( \frac{\partial P}{\partial N}\right )_{T,V} }
     { \left ( \frac{\partial P}{\partial V} \right )_{T,N}} \delta N
 \right ]^2
- \left[ 
 \left ( \frac{\partial \mu}{\partial N} \right )_{T,V} + 
\frac{ \left ( \frac{\partial P}{\partial N}\right )^2_{T,V} }
     { \left ( \frac{\partial P}{\partial V} \right )_{T,N} }
\right ] (\delta N)^2, 
\end{equation}
\end{widetext}
where we have suppressed the suffixes on the right hand side.
Finally from Eqs.\ (\ref{Eq-d2s0}) and (\ref{Eq-d2s2}),
we obtain the stability conditions, 
\begin{eqnarray}
C_v &  > &  0,   \label{Eq-Cv} \\ 
\left ( \frac{\partial P}{\partial V}\right )_{T,N} & < & 0,\label{Eq-pv}   \\
\left ( \frac{\partial \mu}{\partial N} \right )_{T,V}
& > & 
- 
\frac {\left ( \frac{\partial P}{\partial N}\right )^2_{T,V} }
      {\left ( \frac{\partial P}{\partial V} \right )_{T,N} } 
> 0.   \label{Eq-mn}
\end{eqnarray}
We note that (\ref{Eq-mn}) is equivalent to the following
inequality \cite{LaLi},
\begin{equation}\label{Eq-mn2}
\left ( \frac{\partial\mu}{\partial N} \right )_{P,T} >0.
\end{equation}

\end{document}